\documentclass[sigconf,nonacm,balance=false]{acmart}

\newif\ifusecompiledplots
\usecompiledplotstrue

\AtBeginDocument{%
  }

\setcopyright{none}
\copyrightyear{2026}
\acmYear{2026}
\usepackage{amsmath} 
\usepackage{graphicx}
\usepackage{booktabs}
\usepackage{multirow}
\usepackage{xcolor}
\usepackage{colortbl}
\usepackage{subcaption}
\usepackage{float}
\usepackage{tabularx}
\usepackage{array}
\usepackage{textcomp}
\usepackage{algorithm}
\usepackage{algpseudocode}
\usepackage{comment}

\usepackage{xspace}
\newcommand{\name}{\textsc{Scissors}\xspace} 

\newif\ifshowguide
\showguidefalse

\newcommand{\guideblock}[1]{%
  \ifshowguide
    \par\noindent
    {\fbox{\parbox{0.95\linewidth}{NOTES: #1}}}%
    \par
  \fi
}

\newif\ifshowblue
\showbluefalse   
\newif\ifshowgreen
\showgreenfalse

\newcommand{\readytoreview}[1]{%
  \ifshowblue
    {\color{blue}#1}%
  \else
    #1%
  \fi
}

\newcommand{\logan}[1]{%
  \ifshowgreen
    {\color{cyan}#1}%
  \else
    #1%
  \fi
}

\newif\ifshowpehcomments
\showpehcommentsfalse

\newcommand{\pehcomment}[1]{%
  \ifshowpehcomments
    {\bfseries #1}%
  \fi
}

\algnewcommand{\Input}{\item[\textbf{Input:}]}
\algnewcommand{\Output}{\item[\textbf{Output:}]}

\definecolor{aesgray}{RGB}{220,220,220}
\definecolor{suggestionblue}{RGB}{0,80,160}
\excludecomment{suggested}

\begin{document}

\title{Power Side-Channel Evaluation using Multi-FPGA Systems}

\author{Yaswanth Tavva}
\affiliation{%
  \institution{National University of Singapore}
  \country{Singapore}}
\email{yaswanth@u.nus.edu}

\author{Geng Luo}
\affiliation{%
  \institution{National University of Singapore}
  \country{Singapore}}
\email{luo.geng@u.nus.edu}

\author{Trevor E. Carlson}
\affiliation{%
  \institution{National University of Singapore}
  \country{Singapore}}
\email{tcarlson@nus.edu.sg}

\author{Li-Shiuan Peh}
\affiliation{%
  \institution{National University of Singapore}
  \country{Singapore}}
\email{peh@nus.edu.sg}

\renewcommand{\shortauthors}{Tavva et al.}

\begin{abstract}
\readytoreview{Complex out-of-order processors are difficult to evaluate for power side-channel leakage before tapeout. A single FPGA often lacks sufficient capacity, whereas conventional CAD-based analysis is too slow to support the large numbers of traces required by practical attacks. Multi-FPGA emulation provides greater capacity, but existing partitioning schemes are not security-aware and thus results in poor attack performance.

This work presents \name, a security-aware multi-FPGA partitioning approach for pre-silicon power side-channel evaluation of complex out-of-order processors. \name uses side-channel vulnerability factor (SVF) analysis to identify signals correlated with the secret key. It then keeps strongly key-correlated signals confined to one FPGA partition (measured) while moving rest of the signals to the other FPGAs (support), improving the observability of the side-channel information.

We evaluate \name through RTL simulations and physical measurements on a two-U200 FPGA testbed, using frequency-domain CPA with a Hamming-weight leakage model. RTL ablation shows that only 68--71 high-SVF signals dominate key recovery process. For a single-core processor (SingleBOOM), \name recovers the key with 12{,}000 traces, compared with 49{,}000 for the unpartitioned design and 43{,}000 for TritonPart (baseline partitioning method), while improving correct-key correlation by $2.83\times$. For a quad-core processor (QuadBOOM), which exceeds the capacity of the CW340 single-FPGA platform, \name reduces the required traces from 111{,}000 to 66{,}000 and improves correlation by $1.48\times$ over TritonPart. \name also reduces QuadBOOM RTL-to-bitstream generation from 61\,h~39\,min to 6\,h~40\,min, a $9.25\times$ speedup. These results show that security-aware partitioning improves leakage observability and attack performance while substantially reducing implementation time for complex processors against the baseline method.}
\end{abstract}

\begin{suggested}
\paragraph{Suggested text.} Before submission, replace bracketed abstract placeholders with final trace counts, SNR values, correlation values, and overheads. If a value is not final, remove that quantitative claim.
\end{suggested}




\maketitle

\section{Introduction}
\label{sec:intro}

\subsection{Out-of-Order Processors as Side-Channel Targets}

\readytoreview{ Physical side-channel research has traditionally focused on cryptographic accelerators and in-order cores~\cite{balasch2015dpa,longo2015soc,modernsoc2024debao}. Recent studies, however, have demonstrated that complex out-of-order (OoO) processors also exhibit exploitable leakage~\cite{boyapally2025reality,haas2022apple}. Understanding this leakage is increasingly important as OoO processors are adopted in high-assurance and edge-computing systems~\cite{desmet2026masking}.

OoO processors involve many microarchitectural events, such as speculation, dynamic scheduling, physical-register allocation, forwarding, and caching.
These events mix secret-dependent switching with large amounts of \logan{unrelated} background activity.
As a result, leakage from secret computation is harder to observe, and therefore additional traces are required to successfully extract the secret key.
For example, our results show that a SiFive quad-core CPU requires up to 100k traces to extract the unprotected AES-128 key \logan{(Section~\ref{sec:evaluation})}, while prior work on 8-stage~\cite{balasch2015dpa} and 13-stage~\cite{longo2015soc} in-order CPUs could extract the secret within 10k traces.

Existing methods do not support scalable pre-silicon leakage evaluation for large OoO processors. First, software-level analysis does not have visibility into hardware events, and thus cannot capture where secret values influence switching in the hardware. Second, CAD-based simulation exposes switching activity, but it is too slow for trace-heavy attacks on large OoO designs, for the thousands~\cite{boyapally2025reality} to millions~\cite{lipp2021platypus} of traces typically needed. Finally, FPGA emulation provides an alternative evaluation that enables faster execution and the collection of physical measurements, but often cannot fit an OoO processor within a single device.
}

\begin{suggested}
\paragraph{Suggested text.} This paper addresses the gap between scalable processor emulation and realistic side-channel measurement. Single-FPGA side-channel platforms measure small targets well, but cannot host XeonBOOM- and QuadBOOM-class OoO designs. Simulation exposes internal switching activity, but it runs too slowly for trace-heavy attacks or omits measurement noise. \name uses multi-FPGA emulation to study realistic processors while retaining physical power-related measurements.
\end{suggested}

\begin{figure}[t]
  \centering
  \includegraphics[width=\columnwidth]{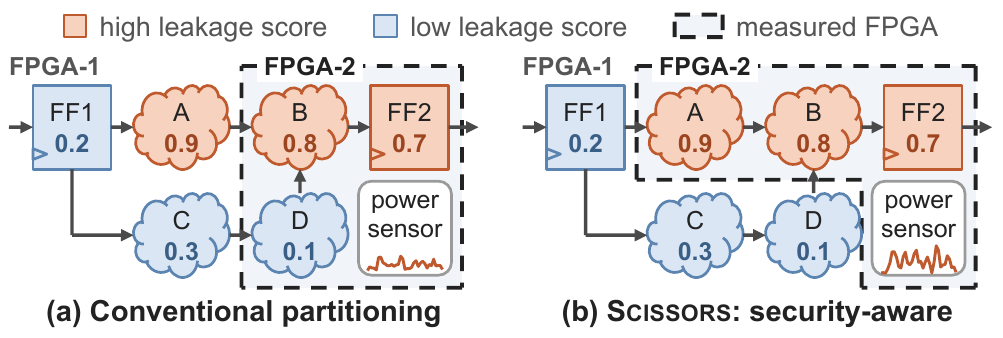}
  \caption{
  Multi-FPGA partitioning of an example circuit.
  (a)~Conventional partitioning cuts for utilization, timing and inter-FPGA
  communication. (b)~\name cuts by SVF score, keeping the high-scoring
  blocks on the measured FPGA.
  }
  \label{fig:overview}
\end{figure}

\subsection{Multi-FPGA Emulation and the Role of Partitioning}

\readytoreview{ Multi-FPGA emulation~\cite{chirkov2023smappic, karandikar2018firesim,prakriya2024tapa,kropotov2026emixemulatingsinglefpgalimits,zhang2025heteroproto,bobda2022future,tessier2015reconfigurable,kapre2009accelerating,dai2017foregraph,zha2020virtualizing,mazraeli2023partitioning} provides a practical path for OoO processors that exceed the capacity of a single device. Partitioning methods such as min-cut and TritonPart~\cite{bustany2023open} enable this approach by addressing FPGA capacity constraints, but they are not designed for side-channel measurements and introduce unintended side effects. For example, they influence logic placement, distribute signals across devices, and determine which signals and logic remain together~\cite{tong2024easypart,zheng2021topopart}. As a result, they unintentionally change how key-dependent and other activity combine in measured traces. Consequently, functionally equivalent partitions can exhibit different leakage characteristics and attack outcomes, either requiring more traces or allowing recovery of only a subset of the key bytes, pointing to the need for security-aware partitioning. \logan{As Figure~\ref{fig:overview}(a) shows, conventional partitioning can leave a block with a high leakage score (A) off the measured FPGA, whereas \name keeps such blocks on the same board (Figure~\ref{fig:overview}(b)).} We present the overview in the subsequent section.}


\subsection{Overview of Approach}

\readytoreview{
We develop \name, a multi-FPGA pre-silicon power side-channel analysis (SCA) evaluation framework of complex designs. Our framework uses FPGA emulation to collect physically measured traces and supports designs that exceed the capacity of single-FPGA platforms.


We decompose the power trace measured on one FPGA into security-relevant
switching activity, unrelated switching activity, and measurement noise:
\begin{equation}
P(t) = P_k(t) + P_o(t) + \eta(t),
\label{eq:power_model}
\end{equation}

where $P_k(t)$ represents activity correlated with the key hypothesis, $P_o(t)$ represents other unrelated microarchitectural activity, and $\eta(t)$ represents measurement noise. CPA attempts to distinguish $P_k(t)$, while $P_o(t)$ and $\eta(t)$ obscure this relationship and increase the number of traces required for key recovery.

Conventional partitioning optimizes resource utilization, timing, and inter-FPGA communication without considering this leakage composition~\cite{karypis1997multilevel,bustany2023open,schlag2023high}.
\logan{It can therefore move key-correlated logic off the measured FPGA, shrinking $P_k(t)$, or keep unrelated logic on it, inflating $P_o(t)$ (Figure~\ref{fig:overview}(a)).} Both
outcomes weaken leakage observability and increase key-recovery effort.


To avoid this degradation, \name incorporates side-channel information directly into the partitioning process. It uses side-channel vulnerability factor (SVF)~\cite{svfisca2012} analysis to rank OoO-core signals according to their relevance to the attack hypothesis.  Our results show that applying the baseline partitioner, TritonPart~\cite{bustany2023open}, can scramble how this activity appears in the measurements, requiring $3.58\times$ as many
traces as \name. Because TritonPart is not security aware and therefore has no knowledge of where these signals should be placed. \name instead retains secret-dependent signals in the measured FPGA power domain and moves weakly correlated activity to the support FPGAs, preserving the observability of
key-dependent activity. Security awareness is therefore necessary because functionally equivalent partitions can expose substantially different mixtures of key-correlated and unrelated switching. Following worst-case side-channel evaluation practice, \name favors leakage observability rather than allowing incidental unrelated activity to be mistaken for security~\cite{azouaoui2020systematic}. We note that \name does not eliminate physical measurement noise or alter processor functionality.

Figure~\ref{fig:bit-hw-hw-svf-rank} shows, unrelated signals alone do not enable key recovery, but they degrade CPA performance when measured together with leakage-relevant signals.

We evaluate \name on XeonBOOM and QuadBOOM (Table~\ref{tab:boom-configs}) as representative complex OoO processors running unprotected AES. The evaluation uses correlation power analysis (CPA), key-rank convergence, measurements to disclosure (MtD), and implementation overhead to quantify trace quality, attack performance, and emulation cost. Overall, \name converts signal-level leakage estimates into security-aware partitioning constraints for RTL designs that require multiple FPGAs. The approach therefore applies to other complex processors, including data-center CPUs, as well as to security-sensitive RTL blocks such as cryptographic accelerators. 
\textbf{This paper makes the following contributions:}

\begin{itemize}
    \item We introduce \name, an SVF-guided, security-aware multi-FPGA partitioning methodology for pre-silicon power-SCA evaluation of complex OoO processors. \name maps leakage-relevant logic to the FPGA that is used for security measurement and distributes the remaining logic across support FPGAs (Figure~\ref{fig:flow}).

    \item We conduct RTL ablation studies on SingleBOOM, QuadBOOM, and XeonBOOM and show that only 68--71 high-SVF signals, out of 43{,}866--174{,}499 signals, dominate key recovery. These signals recover the key within 50 traces in all three configurations; removing them increases SingleBOOM's MtD by $3.3\times$ and prevents recovery for QuadBOOM and XeonBOOM within the trace budget (Figure~\ref{fig:bit-hw-hw-svf-rank} and Table~\ref{tab:svf-selected-signal-counts}).

    \item We evaluate \name on SingleBOOM and QuadBOOM using a two-U200 testbed and use measurements from a HiFive P550 as a physical silicon reference. For SingleBOOM, \name requires 12{,}000 traces, compared with 49{,}000 for the unpartitioned design and 43{,}000 for TritonPart, reducing MtD by up to $4.08\times$ and improving correct-key correlation by up to $2.83\times$. For QuadBOOM, \name reduces MtD from 111{,}000 to 66{,}000 traces ($1.68\times$) and improves correlation by $1.48\times$ (Figures~\ref{fig:quadboom-u200} and ~\ref{fig:singleboom-cpa-comparison}).

    \item We automate the end-to-end SCA workflow and apply it across diverse OoO and in-order CPU configurations (Figure~\ref{fig:frequency-cpa-rank-comparison}). Compared with TritonPart, \name reduces QuadBOOM RTL-to-bitstream time from 61\,h~39\,min to 6\,h~40\,min ($9.25\times$) and XeonBOOM preprocessing time from 16\,h~48\,min to 2\,h~59\,min ($5.61\times$), while reducing the XeonBOOM partition interface from 18{,}946 to 4{,}000 bits (Table~\ref{tab:cpu-partition-cost} and Section~\ref{sec:overhead}).
\end{itemize}

\name also applies to cryptographic accelerators: for an AES hardware implementation, it requires \(4\times\) fewer traces to recover the secret key than the unpartitioned baseline. The unpartitioned experiments use one U200 FPGA, whereas \name uses the two-U200 FPGA testbed. Overall, this work establishes partitioning as a key lever in scalable pre-silicon side-channel evaluation and provides a practical path for analyzing complex processor designs before silicon is available. 

\textbf{Outline}. Section~\ref{sec:background} presents the background, and Section~\ref{sec:multifpga_need} motivates multi-FPGA evaluation. Section~\ref{sec:scissors_method} presents the \name methodology. Sections~\ref{sec:eval_sim} and~\ref{sec:evaluation} report the simulation and physical-measurement results, respectively, while Section~\ref{sec:overhead} evaluates implementation cost and scalability. Sections~\ref{sec:related} and~\ref{sec:discussion} discuss related work and limitations, and Section~\ref{sec:conclusion} concludes.}

\section{Background}
\label{sec:background}
\label{sec:metrics}

\readytoreview{
\textbf{Threat model.} We consider a physical side-channel adversary who chooses inputs, triggers repeated cryptographic executions, and collects power-related traces. We evaluate all attacks in a \textit{known-key setting}, a standard practice that allows us to measure leakage strength, correct-key rank, and key-recovery effort. We focus on software implementation of unprotected AES-128 running on out-of-order processors and use correlation power analysis (CPA) to recover the secret key in a byte-wise
manner.

Measurements use external voltage probes, electromagnetic probes, or on-chip sensors~\cite{drewes2024turn,schellenberg2021inside,quere2026snoopypower,aftabjahani2021special,zhao2018fpga,tian2021remote} that capture voltage fluctuations as a proxy for power activity. Our FPGA setup uses a routing delay sensor (RDS)~\cite{rds2023}, while measurements on the HiFive P550 reference platform use a Keysight oscilloscope~\cite{keysight_msox3104t} and CT4204 Cal Test spring-loaded probe~\cite{caltest_ct4200_series} on the CPU supply rail.

\textbf{Leakage Exploitation through CPA.}
CPA~\cite{brier2004correlation} correlates traces with a hypothetical power model. Let $\mathbf{T}_j=(T_d(j))_{d=1}^{D}$ be the sample $j$ across all traces, and let $\mathbf{H}_i=(HP_i(d))_{d=1}^{D}$ be the hypothetical leakage for key hypothesis $K_i$. We write the CPA correlation as $R(i,j)=\mathrm{corr}(\mathbf{T}_j,\mathbf{H}_i)$, where $\mathrm{corr}(\cdot,\cdot)$ denotes Pearson correlation. The correct hypothesis should produce the largest absolute correlation over $j$. Time-domain CPA considers each sample position across traces, but misalignment among traces can reduce correlation strength.

\textbf{Adaptation to Frequency Domain~\cite{mateos2010new,meynard2010characterization}.}
We transform the traces with an FFT to reduce sensitivity to trace misalignment. Let $\mathbf{F}_{j'}=(F_d(j'))_{d=1}^{D}$ denote the FFT magnitude at frequency index $j'$ across all traces. We reuse the same correlation operator and compute $R(i,j')=\mathrm{corr}(\mathbf{F}_{j'},\mathbf{H}_i)$ across $N_{\mathrm{FFT}}/2$ frequency components. We use frequency domain CPA in this work.

\textbf{Security Metrics.}
We evaluate CPA with score vector $\mathbf{p}$, where $p_i$ is the maximum absolute correlation for key hypothesis $K_i$ over the relevant time or frequency components. We use key rank and measurements to disclosure (MtD).

\textbf{Key Rank.}
Let $i^\star$ denote the hypothesis index of the correct key $K_{i^\star}$. The key rank is:
\begin{equation}
\mathrm{Rank}(K_{i^\star})
= 1 + \left|\{i \mid p_i > p_{i^\star}\}\right|.
\end{equation}

\textbf{Measurements to Disclosure (MtD).}
MtD is the minimum trace count to achieve $\mathrm{Rank}(K_{i^\star})=1$.

\textbf{Side-channel Vulnerability Factor (SVF)~\cite{svfisca2012}.}
SVF measures the correlation between internal signal activity and an attack hypothesis. For signal value $x_d[t]$ and hypothesis value $h_d$ in execution $d$, we define
\begin{equation}
\begin{gathered}
a_d[t] = \mathrm{HD}(x_d[t],x_{d+1}[t]), \qquad
o_d = \mathrm{HD}(h_d,h_{d+1}), \\
\mathrm{SVF}(x)
= \max_{t\in\{1,\ldots,N_t\}}
\left|\mathrm{corr}\big((o_d)_d,(a_d[t])_d\big)\right|.
\end{gathered}
\end{equation}
Here, $\mathrm{HD}$ denotes Hamming distance, and the SVF score reuses the Pearson correlation operator defined above.
Thus, a high SVF score indicates that a signal's switching activity closely follows the attack hypothesis and is therefore more likely to contribute to successful key recovery.

\textbf{Toggle-count Based Simulation~\cite{veshchikov2017use}.}
PrimePower and SPICE-based simulators provide high-fidelity power
waveforms, but they are difficult to scale due to long runtime and cost licensing. We therefore use bit-normalized Hamming-weight activity traces directly from the
VCD signal states, as a proxy for power consumption. Let $\mathcal{S}$ denote the selected design signals
and $w_s$ the bit-width of signal $s$. The normalized activity at time
sample $t$ is
\begin{align}
A_{\mathrm{HW}}^{\mathrm{norm}}[t]
=
\frac{
    \sum_{s \in \mathcal{S}} \mathrm{HW}\bigl(s[t]\bigr)
}{
    \sum_{s \in \mathcal{S}} w_s
}.
\end{align}

 The resulting traces are correlated against the AES
S-box Hamming-weight hypothesis for each key guess. We list the notations in Table~\ref{tab:terminology-notation}. For more elaborate discussion on each metric refer to ~\cite{papagiannopoulos2023side}.

}

\begin{table}[H]
\centering
\caption{Terminology and Notation.}
\label{tab:terminology-notation}
\scriptsize
\setlength{\tabcolsep}{4pt}
\renewcommand{\arraystretch}{0.96}
\begin{tabularx}{\columnwidth}{@{}lX@{}}
\toprule
\textbf{Term or notation} & \textbf{Meaning} \\
\midrule

\multicolumn{2}{@{}l}{\textit{Power and CPA notation}} \\
$P(t)$ & Power-related trace observed by the sensor at time $t$. \\
$P_k(t)$, $P_o(t)$, $\eta(t)$ & Key-correlated activity, other microarchitectural activity, and measurement noise, respectively. \\
$D$ & Number of traces in the CPA equations; in Algorithm~\ref{alg:scissors_partition}, the input RTL design. \\
$d$ & Trace or execution index in CPA and SVF; request or response phase index in the communication model, where $d\in\{\mathrm{req},\mathrm{resp}\}$. \\
$i$, $i^\star$ & Key-hypothesis index and the correct-key hypothesis index, respectively. \\
$j$, $j'$ & Time-domain sample index and frequency-domain index, respectively. \\
$t$ & Time/sample index in the power model and RTL simulation. \\
$T_d(j)$, $\mathbf{T}_j$ & Sample $j$ of measured trace $d$ and the vector of sample $j$ across all $D$ traces, respectively. \\
$K_i$, $K_{i^\star}$ & Candidate key represented by hypothesis index $i$ and the correct key, respectively. \\
$HP_i(d)$, $\mathbf{H}_i$ & Hypothetical leakage for key hypothesis $K_i$ in trace $d$ and the corresponding vector across all $D$ traces, respectively. \\
$\mathrm{corr}(\cdot,\cdot)$ & Pearson correlation operator used by CPA and SVF. \\
$R(i,j)$, $R(i,j')$ & Pearson correlation for hypothesis $i$ at time-domain sample $j$ or frequency-domain index $j'$, respectively. \\
$F_d(j')$, $\mathbf{F}_{j'}$ & FFT magnitude of trace $d$ at frequency index $j'$ and the corresponding vector across all $D$ traces, respectively. \\
$N_{\mathrm{FFT}}$ & FFT length; a real-valued trace provides $N_{\mathrm{FFT}}/2$ analyzed frequency components. \\
$\mathbf{p}$, $p_i$ & CPA score vector and the maximum absolute correlation for hypothesis $i$ over the relevant time or frequency components. \\
$\mathrm{Rank}(K_{i^\star})$ & One-based rank of the correct key according to the CPA scores; rank~1 denotes the top candidate. \\
$\rho$ & Correlation coefficient; $|\rho|$ denotes its absolute value. \\

\midrule
\multicolumn{2}{@{}l}{\textit{SVF and RTL-simulation notation}} \\
$x_d[t]$, $h_d$ & Value of an RTL signal at sample $t$ and the corresponding hypothesis value in execution $d$. \\
$a_d[t]$, $o_d$ & Hamming-distance activity of the RTL signal and the leakage oracle, respectively. \\
$\mathrm{HD}(\cdot,\cdot)$, $\mathrm{HW}(\cdot)$ & Hamming-distance and Hamming-weight operators, respectively. \\
$\mathrm{SVF}(x)$, $|\rho_{\mathrm{SVF}}|$ & Maximum absolute correlation assigned to signal $x$ and the equivalent SVF-correlation magnitude used for signal selection. \\
$N_t$ & Number of time samples considered in the SVF maximization. \\
$\mathcal S$, $s$, $w_s$ & Selected signal set, a signal in that set, and its bit width. \\
$A_{\mathrm{HW}}^{\mathrm{norm}}[t]$ & Bit-normalized aggregate Hamming-weight activity at sample $t$. \\
$N$ & Number of aligned workload windows extracted from the VCD. \\
$\tau$, $\mathcal S_\tau$ & SVF threshold and the set of signals that satisfy it. \\

\midrule
\multicolumn{2}{@{}l}{\textit{Partitioning and communication notation}} \\
$\mathcal B$, $\mathcal X$ & Sets of black-box descendants and excluded instances, respectively. \\
$\mathcal I_\tau$, $I$ & Deduplicated set of selected RTL instances and one instance in that set. \\
$P_0$, $P_1$ & Support and measured RTL partitions, respectively. $P_0$ contains the remaining support logic and is implemented on the support FPGA, while $P_1$ contains the selected instances and is implemented on the sensor-equipped measured FPGA. \\
$K$ & Number of FPGAs in the multi-FPGA system. \\
$\Gamma$ & Partition-boundary map containing crossing-signal directions and widths. \\
$B_{uv}$, $B_{01}$, $B_{10}$ & Functional boundary width from partition $u$ to $v$, including the two directions of a two-FPGA partition. \\
$p$, $q_{uv}$, $F$ & Functional bits per packet, number of packets from $u$ to $v$, and physical packet width. \\
$e$, $b$ & Physical-link index and the number of physical bits transferred on that link. \\
$L_e(b)$, $L_{0,e}$, $R_{\mathrm{eff},e}$ & Link-$e$ latency for $b$ bits, its fixed latency, and its effective data rate. The multi-FPGA model uses the equivalent forms $L_{0,\ell}$ and $R_{\mathrm{eff},\ell}$ when indexing links by $\ell$. \\
$T_{\mathrm{upd}}$ & Two-FPGA request--response update time. \\
$U_{\max}$ & Maximum number of cross-partition settling updates within one logical cycle. \\
$T_{\mathrm{emu}}$, $f_{\mathrm{emu}}$ & Emulation-clock period and its maximum safe frequency. \\
$\mathcal E_L$, $\ell$ & Set of physical links and a link in that set, respectively. \\
$r$, $n_{\ell,d}^{(r)}$, $b_{\ell,d}^{(r)}$ & Request--response round index, number of messages, and physical-bit count on link $\ell$ during the request or response phase $d\in\{\mathrm{req},\mathrm{resp}\}$ of round $r$. \\
$T^{(r)}$ & Duration of request--response round $r$. \\
\bottomrule
\end{tabularx}
\end{table}

\section{Limitations of Existing Pre-Silicon Approaches}
\label{sec:multifpga_need}

\begin{figure}[t]
    \centering
    \resizebox{0.75\linewidth}{!}{%
    \includegraphics[width=\linewidth]{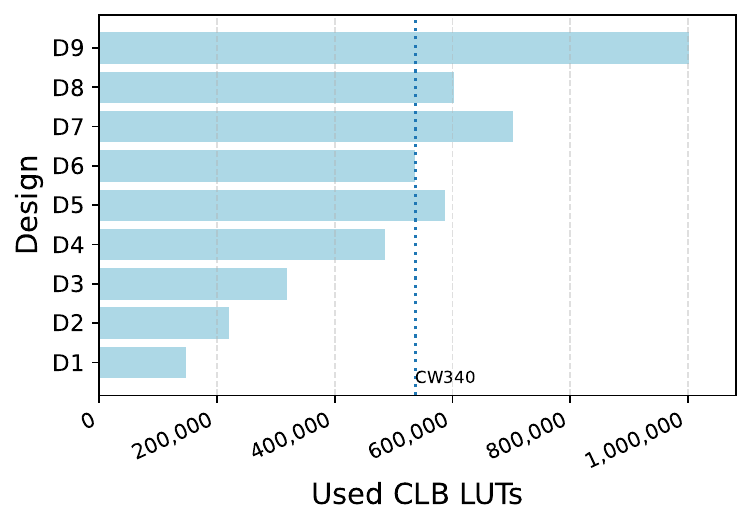}
    }
    \caption{LUT usage across BOOM variants \logan{D1--D9}. Dotted blue line marks the capacity of the CW340/KU095 board.}
    \label{fig:platform_fit}
\end{figure}

\readytoreview{Existing pre-silicon SCA methods do not simultaneously provide sufficient simulation speed, realistic physical measurement conditions, and the capacity required for complex out-of-order (OoO) processors. Gate-level simulation provides detailed switching activity~\cite{kiaei2021saidoyoki}, but it is inherently noiseless. Moreover, simulations of BOOM-based OoO processors run at only a few processor cycles per second~\cite{kim2016strober}, making the collection of the large trace sets required for practical attacks prohibitively slow. SPICE-level noise-injection methods provide more detailed physical models for small designs~\cite{anik2024resiliency}, but their computational makes it difficult to use on large designs like an OoO processor due to long runtimes.
\logan{\name uses RTL simulation only to rank signals, which needs far fewer runs than the tens of thousands of traces a CPA attack requires.}

FPGA emulation substantially increases execution throughput and enables power measurements under physical operating conditions. Although FPGA lookup tables, routing resources, and power-delivery networks differ from ASIC standard cells and interconnects, both implement the same RTL-level computations. FPGA measurements can therefore reveal whether these computations produce statistically exploitable leakage, even when the precise leakage magnitude, timing, and spatial distribution may differ from those of an ASIC~\cite{tavva2025ctscan}. Platforms such as ChipWhisperer~\cite{o2014chipwhisperer}, SAKURA~\cite{guntur2014side}, FOBOS~\cite{ferrufino2023fobos}, and the OpenTitan CW310~\cite{cw310} and CW340 boards~\cite{cw340} provide established measurement infrastructures for FPGA-based SCA.
\logan{Each of them hosts the target design on a single FPGA.}

The capacity of a single FPGA, however, limits this approach for complex OoO designs.
As Figure~\ref{fig:platform_fit} shows, OoO configurations \logan{(Table~\ref{tab:boom-configs})} can exceed the capacity of the largest available side-channel analysis board, the CW340 \logan{(Table~\ref{tab:platforms})}.
In particular, configuration D9 (Table~\ref{tab:boom-configs}),
\logan{the closest match to the quad-core HiFive P550 we use as reference,}
requires about 934k LUTs, 73\% more resources than the CW340 can support.
Increasing the issue width, reorder-buffer size, cache capacity, or core count can produce designs that exceed even an Alveo U200~\cite{amd_alveo_u200}.
\logan{The XeonBOOM configuration we evaluate, is one such design.}
Multi-FPGA emulation addresses this capacity limitation by distributing the complete design across multiple devices while retaining FPGA execution speed and physical power measurement.}


\begin{table}[ht]
\centering
\caption{Resource availability of platforms.}
\label{tab:platforms}
\resizebox{\linewidth}{!}{
\begin{tabular}{|l|c|r|r|r|r|c|}
\hline
\textbf{\shortstack{FPGA\\Platform}} &
\textbf{\shortstack{FPGA\\Device}} &
\multicolumn{1}{c|}{\textbf{\shortstack{Look-Up Tables\\(LUTs) (k)}}} &
\multicolumn{1}{c|}{\textbf{\shortstack{Registers\\(k)}}} &
\multicolumn{1}{c|}{\textbf{\shortstack{Block RAMs\\(Mb)}}} &
\multicolumn{1}{c|}{\textbf{\shortstack{DSP\\Slices}}} &
\textbf{\shortstack{Measurement\\Type}} \\ \hline
CW340~\cite{cw340,amd_ultrascale_fpga_selection_guide_2016} &
XCKU095 & 537 & 1075 & 59 & 768 & Probe \\ \hline
Alveo U200~\cite{amd_alveo_u200} &
XCU200 & 1182 & 1956 & 67 & 5880 & on-chip sensor \\ \hline
\end{tabular}
}
\end{table}

\begin{table}[ht]
\centering
\caption{Microarchitectural configurations for the design points shown in \logan{Figure~\ref{fig:platform_fit}}. D10~\cite{whangbo2024fireaxe}, D11~\cite{lam2024lioncove}, D12~\cite{johnson_firestorm} are literature-derived survey points and are not the FPGA implementations.
}
\smallskip
\parbox{\linewidth}{\footnotesize
D1: Rocket64w1;
D2: Rocket64x1;
D3: SingleBOOM;
D4: Rocket64z1;
D5: Rocket64c9101;
D6: Rocket64g1;
D7: Rocket64gc401;
D8: Rocket64x4;
D9: QuadBOOM;
D10: XeonBOOM;
D11: Lion Cove-like; and
D12: Apple Firestorm-like.
}

\resizebox{0.99\linewidth}{!}{
\begin{tabular}{|l|c|c|c|c|c|c|c|c|c|c|c|c|}

\hline
\textbf{Parameter}
& \textbf{D1} & \textbf{D2} & \textbf{D3} & \textbf{D4}
& \textbf{D5} & \textbf{D6} & \textbf{D7} & \textbf{D8}
& \textbf{D9} & \textbf{D10} & \textbf{D11} & \textbf{D12} \\ \hline

Cores
& 1 & 1 & 1 & 1 & 1 & 1 & 1 & 4 & 4 & 1 & 1 & 1 \\ \hline

Decode width
& 1 & 2 & 3 & 4 & 4 & 5 & 6 & 2 & 3 & 6 & 8 & 8 \\ \hline

ROB entries
& 32 & 64 & 96 & 128 & 192 & 130 & 216 & 64 & 96 & 510
& 576 & $\sim$623 \\ \hline

I-Phys Regs
& 52 & 80 & 100 & 128 & 160 & 128 & 115 & 80 & 100 & 280
& $\sim$290 & $\sim$380--394 \\ \hline

F-Phys Regs
& 48 & 64 & 96 & 128 & 160 & 128 & 132 & 64 & 96 & 332
& $\sim$406 & $\sim$432 \\ \hline

Ld queue entries
& 8 & 16 & 24 & 32 & 40 & 32 & 76 & 16 & 24 & 192
& $\sim$189 & $\sim$130 \\ \hline

St queue entries
& 8 & 16 & 24 & 32 & 40 & 32 & 45 & 16 & 24 & 114
& 120 & $\sim$60 \\ \hline

Fetch buffer entries
& 8 & 16 & 24 & 32 & 48 & 35 & 54 & 16 & 24 & 144
& 192$^\dagger$ & -- \\ \hline

L1-I
& 16 kB & 16 kB & 32 kB & 32 kB & 64 kB & 32 kB
& 32 kB & 16 kB & 32 kB & 32 kB & 64 kB & 192 kB \\ \hline

L1-D
& 16 kB & 16 kB & 32 kB & 32 kB & 64 kB & 32 kB
& 32 kB & 16 kB & 32 kB & 32 kB
& 48 kB  & 128 kB \\ \hline

\end{tabular}}
\label{tab:boom-configs}
\end{table}

\begin{figure*}[t]
\centering

\begin{minipage}[t]{0.9\textwidth}
  \centering
  \includegraphics[width=\linewidth]{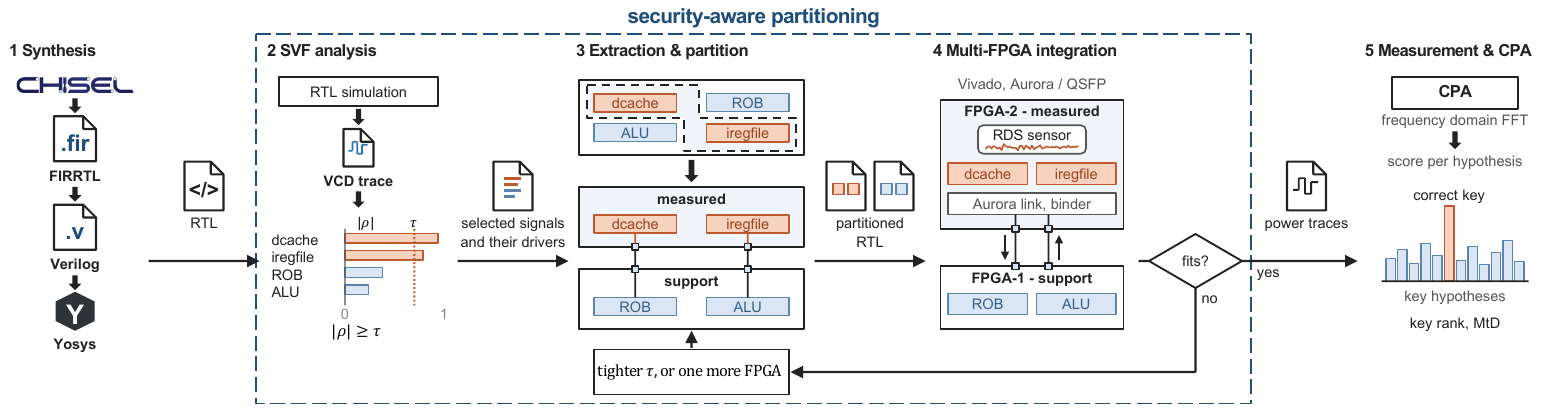}
\end{minipage}

\caption{
\logan{Overview of \name.
}
}
\label{fig:flow}
\end{figure*}

\section{\name: SVF-Guided Security-Aware Partitioning}
\label{sec:scissors_method}

\name uses an RTL design, a representative workload, a leakage model, an
attack hypothesis, and a target FPGA platform. It simulates the workload at
RTL level and ranks signals by their Side-channel Vulnerability Factor (SVF) (Section ~\ref{sec:background}). These
scores guide the partitioning process shown in Figure~\ref{fig:flow}. \name
first identifies the RTL instances that contain strongly key-correlated
signals and anchors them to the \emph{measured FPGA}. It then assigns the
remaining hierarchy to the \emph{support FPGAs}.
In this way, the logic most likely to expose the secret remains concentrated
on the measured FPGA, while the remaining logic is moved to the support FPGAs,
thereby increasing the relative contribution of key-related leakage, i.e., $P_k(t)$. Equation~\eqref{eq:power_model} captures this effect.
Section~\ref{sec:eval_sim} explains why the \name approach works, and
Section~\ref{sec:evaluation} demonstrates it through FPGA
measurements. We detail each phase of our flow in the following subsections.


\subsection{{Scoring Signals and Selecting RTL Instances}}
\label{sec:scissors_svf}

The testbench records RTL activity and workload-iteration markers in a
value-change dump (VCD). \name divides the trace into $N$ workload-aligned
windows, computes the SVF metric in Section~\ref{sec:metrics}, and collects the
signals whose scores meet the threshold $\tau$ in $\mathcal S_\tau$. Each run
records the simulated hierarchy, leakage oracle, SVF scores, SVF threshold,
and the signal set associated with that threshold.

{The VCD captures the signal activity used to estimate leakage during simulation. Simulation allows us to selectively include signals in the analysis, i.e., focus on a specific signal set, and exclude the remaining signals. In
practice, this can only be achieved by moving the signals physically across FPGAs and
placing them on the FPGA of our choice. \name therefore maps every signal in $\mathcal S_\tau$
to the exact enclosing RTL instance that implements it. In other words, selection is based on an instance's hierarchical
occurrence; thus, selecting one instance does not select
other instances of the same module in another core. \name discards candidates
inside memory macros, existing black boxes, or their descendants because their
synthesizable RTL is unavailable. The visible logic around those macros
remains eligible, and the macro declarations are retained for synthesis. We next detail the process for moving the selected RTL instances.}

\subsection{{Moving Selected Instances and Validating Each Boundary}}
\label{sec:scissors_extraction}

\begin{figure}[t]
  \centering
  \includegraphics[width=\columnwidth]{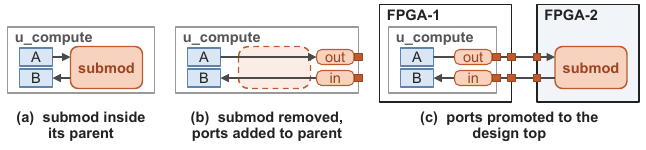}
  \caption{
  Extracting source logic. (a)~The driver cells of the selected
  signals, grouped into a submodule inside their source instance. (b)~Removing
  the submodule reverses each connection at the parent: a submodule input
  becomes a parent output. (c)~Ports are added up to the design top, which links
  to the submodule on the measured FPGA.
  }
  \label{fig:extract}
\end{figure}

{Using the resulting instance set, \name moves each selected
hierarchical occurrence exactly once. The complete RTL hierarchy rooted at
that occurrence becomes part of the measured partition. If
the selected module type is reused elsewhere, \name specializes only the
modules along the selected occurrence's path to the design top, leaving all
other occurrences unchanged.

Detaching an instance exposes its existing interface as the partition
boundary. A child input becomes an output of the support partition and an
input of the measured partition, while a child output crosses in the opposite
direction. \name lifts these connections through the specialized ancestors
and records every signal, direction, and width in the boundary map $\Gamma$.
The selected instances from Step~\ref{sec:scissors_svf} are moved instances to form the measured partition $P_1$, and the unchanged remainder
forms the support partition $P_0$. Local clocks are used. Figure~\ref{fig:extract} illustrates this process. Finally, a temporary top
directly reconnects $P_0$ and $P_1$; If more than one instance to be moved, \name accepts each incremental boundary
only when the original application-level test still completes.}
Algorithm~\ref{alg:scissors_partition} summarizes this construction. $P_0$ and $P_1$ are partitions of the same design and are implemented on two separate FPGAs. This can be generalized across any number of support partitions, as described in Section~\ref{sec:scissors_scaling}. Our next section describes the binder that enables seamless communication between the FPGAs. 

\begin{algorithm}[t]
\caption{SVF-Guided Per-Instance Partition Construction}
\label{alg:scissors_partition}
\begin{algorithmic}[1]
\Input RTL $D$, selected signals $\mathcal S_\tau$, black boxes and descendants
$\mathcal B$, excluded instances $\mathcal X$
\Output Support RTL $P_0$, measurement RTL $P_1$, boundary map $\Gamma$
\State {Discard signals within $\mathcal B$ or $\mathcal X$}
\State {$\mathcal I_\tau\gets$ unique hierarchical RTL instances
containing the remaining selected signals}
\For{{each selected instance $I\in\mathcal I_\tau$}}
  \State {Clone modules along the hierarchical path from $I$ to the design top}
  \State {Detach the complete instance $I$ while preserving its RTL hierarchy}
  \State {Lift the input and output connections of $I$ to top-level ports}
  \State {Update $\Gamma$ with the direction and width of each crossing signal}
  \State {Reconnect the partitions in a temporary top; reject the
  boundary if validation fails}
\EndFor
\State {Aggregate $\mathcal I_\tau$ into $P_1$; set $P_0$ to the
remaining hierarchy; emit $\Gamma$}
\State Keep clocks local; \Return $P_0,P_1,\Gamma$
\end{algorithmic}
\end{algorithm}


\subsection{Binder, Resource Closure, and Frequency}
\label{sec:scissors_binder}

{The accepted partitions and boundary map $\Gamma$ determine
what the two FPGAs must exchange to emulate one synchronous design.} For each
direction, an adapter packs the crossing signals into a $B_{01}$- or
$B_{10}$-bit vector. A \emph{binder} packetizes these vectors for the
finite-width physical links~\cite{babb1993virtual}, transmits them, and
reassembles the original vectors at the receiver. Packet tags identify the
partition, logical cycle, exchange number, vector position, and completion
status.


A cut can create a cross-partition dependency chain within one logical cycle.
$P_0$ sends inputs to $P_1$, and the outputs produced by $P_1$ return to
$P_0$. Re-evaluating $P_0$ may then produce another input for $P_1$. Because
each exchange depends on the result of the previous one, the binder performs
these exchanges sequentially. \textit{This behavior is common to multi-FPGA
partitioning and applies to both \name and TritonPart.} In the best case, no
combinational dependency crosses the partition boundary. Both directions then
transfer concurrently, so the slower one-way transfer determines the latency.
However, a combinational path that crosses the boundary multiple times may
require up to $U_{\max}$\footnote{We determine $U_{\max}$ immediately after
partitioning by statically tracing combinational dependency chains across
partition boundaries} updates. We therefore select $T_{\mathrm{emu}}$ to provide
enough time for all exchanges to complete and all crossing values to settle
before the next logical clock edge. Consequently, the boundary width and
communication latency determine the highest safe emulation frequency.

\pehcomment{Peh: Explain why you need to estimate communication time. Do u try to reduce communication time after putting the SVF parts on fpga 1? I thought it's just inevitable that Scissors will have more communication, and it'll show in the results, but need not be modeled? {\color{purple} yash: added the just justification for estimating the communication time. we need some notion of what is a practical frequency value to try. }}
{
\name does not use communication cost to select the measured partition. SVF
makes this selection first. The communication model then determines a safe
clock period for its dependent
exchanges to complete within one logical cycle.
}
If each packet carries $p$ functional bits, transferring the boundary in
direction $(u,v)$ requires $q_{uv}=\lceil B_{uv}/p\rceil$ packets. \name
characterizes physical link $e$ with
\begin{equation}
 L_e(b)=L_{0,e}+b/R_{\mathrm{eff},e},
 \label{eq:scissors_link_model}
\end{equation}
where $b$ includes packet headers and padding, $L_{0,e}$ is the fixed link
latency, and $R_{\mathrm{eff},e}$ is the effective data rate. \pehcomment{Peh: Point to Fig 5 and briefly explain with the concrete example of 40Gb/s link here. Then continue the walkthrough later all the way to Temu, with specific numbers of packet width, etc. Else reviewers won't have the patience to follow. {\color{purple} yash: added the example.}}


{
Consider the 40-Gb/s QSFP FPGA to FPGA link~\cite{xilinx_aurora_alveo_vitis_2022}. Each 256-bit physical packet carries $p=208$
functional bits and 48 metadata bits. The QuadBOOM \name boundary contains 4{,}013
bits from $P_0$ to $P_1$ and 2{,}397 bits in the reverse direction
(Table~\ref{tab:cpu-partition-cost}). The binder therefore sends
$q_{01}=\lceil4{,}013/208\rceil=20$ packets and
$q_{10}=\lceil2{,}397/208\rceil=12$ packets. Including headers and padding,
these transfers contain $20\times256=5{,}120$ and
$12\times256=3{,}072$ physical bits. 
}


Using physical packet width $F$, one two-FPGA update takes
\begin{equation}
 T_{\mathrm{upd}}=L_{01}(q_{01}F)+L_{10}(q_{10}F).
 \label{eq:scissors_update_time}
\end{equation}
Because the returned vector depends on the request, a full-duplex link cannot
eliminate this ordered sequence. The logical emulation clock must satisfy
\begin{equation}
 T_{\mathrm{emu}}\geq U_{\max}T_{\mathrm{upd}},
 \qquad f_{\mathrm{emu}}\leq 1/T_{\mathrm{emu}}.
 \label{eq:scissors_frequency}
\end{equation}

As an example, let's look at the requirements for a QuadBOOM implementation. Substituting
the packet counts above into Equation~\eqref{eq:scissors_update_time} gives
\[
 T_{\mathrm{upd}}=L_{01}(5{,}120)+L_{10}(3{,}072).
\]
With the measured $L_{0,e}=380$~ns and $R_{\mathrm{eff},e}=39.94$~Gb/s, the
two directional transfers take 508.19 and 456.92~ns, respectively, giving
$T_{\mathrm{upd}}=965.11$~ns. Our binder permits at most $U_{\max}=5$
dependent updates per logical cycle, so $T_{\mathrm{emu}}\geq4.83~\mu$s and
$f_{\mathrm{emu}}\leq207.23$~kHz. Our implementation conservatively uses a
100-kHz processor clock, corresponding to $T_{\mathrm{emu}}=10~\mu$s. A
mapping that exceeds this period must run at a lower frequency.


Meeting the communication bound is necessary but not sufficient because each
partition must also fit within its device. Section~\ref{sec:scissors_scaling} next
describes scaling beyond two FPGAs and applies to both \name
and TritonPart. 


\subsection{Scaling Beyond Two FPGAs}
\label{sec:scissors_scaling}

{Even beyond two FPGAs, \name's security decision remains the
same: all selected instances stay together on one measured device. \name places
only the support logic across the remaining $K-1$ devices, subject to resource constraints, i.e., Step~\ref{sec:scissors_extraction} is recursively
applied to create new partitions. Also, adding support FPGAs cannot make
an oversized measured partition fit. That requires changing the security cut,
for example, by raising $\tau$ and omitting lower-SVF instances. Likewise, if an $k$-way TritonPart partition does not fit on $K$ FPGAs, we incrementally increase the number of partitions and FPGAs.}

Fitting the support logic by capacity is straightforward; the main challenge is determining how much
time all cross-partition combinational paths require to settle before the next
logical clock edge. We use the $T_{\mathrm{emu}}$ model to select a
sufficiently long logical-clock period. During this period, all FPGAs repeatedly
exchange their current boundary values and evaluate their local combinational
logic.

We trace each combinational path across
partition boundaries while treating registers and memories as endpoints. The
maximum number of FPGA boundaries crossed by any such path determines the
number of settling rounds required within one logical-clock period. In each
round, the FPGAs exchange their current boundary values, after which their
local combinational logic responds to the newly received values. The logical
clock period must accommodate enough rounds for every path to settle at a
register or memory input before the next clock edge.

If link $\ell$ carries $n_{\ell,d}^{(r)}$ messages totaling
$b_{\ell,d}^{(r)}$ bits during phase
$d\in\{\mathrm{req},\mathrm{resp}\}$ of request--response round $r$, the
duration of that round is bounded by
\begin{equation}
 T^{(r)}
 =
 \sum_{d\in\{\mathrm{req},\mathrm{resp}\}}
 \max_{\ell\in\mathcal E_L}
 \left(
 n_{\ell,d}^{(r)}L_{0,\ell}
 +
 \frac{b_{\ell,d}^{(r)}}{R_{\mathrm{eff},\ell}}
 \right).
 \label{eq:scissors_multifpga_round}
\end{equation}





We sum $T^{(r)}$ over at most $U_{\max}$ settling rounds to determine a safe
logical-clock period, just as in the two-FPGA case. Application-level simulation and physical timing closure
provide the final acceptance tests.

\pehcomment{Peh: Does Scissors search the design space for partition that reaches the fastest emulation clock? Given that there are only 3 or 4 boards, and the boards are identical with just one link between boards, isn't it fairly straightforward to fit as much of the non-secure-aware parts as possible onto the 3rd board, and the leftover on the 4th? :). Explain better what this Temu estimate is for, and how it is used in the partitioning, and how it differs from prior FPGA partitioning algos {\color{purple} yes. capacity very straightforward. more challenging is if partitions have a long cross-partition Combinational logic chain. not a unique problem to scissors. we also see in tritonpart. tritonpart tires not to make the cut on the critical path, but still not a guranteee that resultant cut has no such back and forth Combinational paths between partitions across fpgas. -- text is almost there. need some more polish.}}


\subsection{Timing-Aware Baseline}
\label{sec:scissors_tritonpart}

{To isolate the benefit of the security-aware cut, we compare
against TritonPart, OpenROAD's timing-aware hypergraph partitioner. TritonPart
partitions synthesized cells and shared-net hyperedges to satisfy timing and
balance constraints; it does not model SVF scores or RTL instance paths. Both methods
start from the same RTL and use the same black-box policy, FPGA count, and
platform constraints. The distinction is the movable unit and the objective:
TritonPart cuts cells to optimize a physical cost, whereas \name moves exact
RTL instances selected for their key-correlated activity, as described in
Sections~\ref{sec:scissors_svf} and~\ref{sec:scissors_extraction}.}

\pehcomment{Peh: Doesn't TritonPart also need to estimate communication time? Basically unclear up to this point of the paper why 4.2, 4.3, 4.4 of Scissors is unique to Scissors. Do explain better the differences (other than 4.1) in steps 4.2-4.4 from TritonPart {\color{purple} yes. 4.1 and 4.2 for scissors alone. 4.3 and 4.4 are not unique to scissors. binder must work for both scissors and tritonpart. i re wrote the whole section 4. need some more polishing.}}

{
The binder, communication model, resource checks, and functional validation
in Section~\ref{sec:scissors_binder}, along with the scaling method in
Section~\ref{sec:scissors_scaling}, form a common realization layer for both
TritonPart and \name. We apply this layer identically to both methods so that
the comparison isolates the boundary-selection policy.
}

\subsection{Execution Flow}

After both partitions pass functional and feasibility checks, we use Vivado to generate their bitstreams. Then, we program the FPGAs, initialize the
links, and calibrate the on-chip sensor on the measured FPGA. The acquisition
controller relates the sensor's 100-MHz sampling clock to the emulated design
clock, sizes the BRAM recording window, and uses software markers to delimit
AES-128 S-box execution. During an encryption, the binder performs the boundary
updates required for each logical cycle while the sensor records fine-grained
propagation-delay variations as a proxy for power consumption. At the end of the marked
region, the measured FPGA streams the buffered trace to the host and reports
completion; software can then begin the next encryption. These traces are later used to launch the CPA attack.


\section{Simulation Evaluation}
\label{sec:eval_sim}

\readytoreview{This section uses RTL simulation to determine how strongly high-SVF signals influence CPA-based key recovery. RTL simulation provides visibility into individual signal values, allowing us to choose selected signal groups and measure their effect on key rank and measurements to disclosure (MtD). We first evaluate the high-SVF signals alone, then combine them with all selected signals, and finally remove them to quantify their contribution across XeonBOOM, QuadBOOM, and SingleBOOM.

For this experiment, each processor configuration executes an unprotected software implementation of AES. We perform CPA using a Hamming-weight leakage model and rank the key candidates according to their correlation with the simulated RTL activity. 

To perform the ablation study, we divide the signals into the SVF-correlation ranges listed in Table~\ref{tab:svf-selected-signal-counts}. Although each configuration contains tens of thousands of selected signals, fewer than 100 have $|\rho_{\mathrm{SVF}}| \geq 0.9$. Figure~\ref{fig:bit-hw-hw-svf-rank} shows that this small high-SVF subset dominates key recovery.

We first perform CPA using only signals with $|\rho_{\mathrm{SVF}}| \geq 0.9$. CPA recovers the correct key within 50 traces for all three configurations, demonstrating that these signals contain highly concentrated key-dependent leakage. We next include all selected signals to examine how weaker signals affect the attack. CPA still recovers the key for SingleBOOM and QuadBOOM, but requires approximately $6\times$ and $16\times$ more traces, respectively, than the high-SVF-only experiment. For XeonBOOM, the correct key was not extracted, although it remains among the top three candidates. Thus, combining the high-SVF signals with a much larger set of weaker signals preserves useful leakage but reduces attack performance.

To isolate the importance of the high-SVF subset, we then exclude all signals with $|\rho_{\mathrm{SVF}}| \geq 0.9$. This removal has a substantial effect despite eliminating only 68--71 signals from each configuration. SingleBOOM requires nearly 700 additional traces to recover the key, corresponding to a $3.3\times$ increase. More importantly, CPA no longer recovers the key for QuadBOOM or XeonBOOM within the trace budget.

Further restricting the signal set strengthens this trend. When we retain only signals with $|\rho_{\mathrm{SVF}}| < 0.5$, the correct-key rank degrades substantially for SingleBOOM, while QuadBOOM and XeonBOOM continue to resist key recovery. The results therefore show that lower-SVF signals cannot compensate for the removal of the small high-SVF subset.

Overall, high-SVF signals provide most of the leakage that enables CPA-based key recovery. These signals reveal the key efficiently when analyzed alone and continue to support recovery when combined with weaker signals, although the weaker signals increase MtD. In contrast, removing the high-SVF signals prevents key recovery in two configurations and degrades it in the third.

This RTL study covers the TileGroup, which includes the processor core and L1 cache. The FPGA implementations evaluated in the next section cover the complete SoC, including the L2 cache, DDR controller, and peripheral subsystems. These findings guide our security-aware partitioning strategy, which confines the placement of high-SVF signals to one particular FPGA, while lower-SVF activity is moved away to the support FPGAs.}

\begin{suggested}
\paragraph{Suggested text.} Before the final pass, make the SVF setup consistent. The text says SVF uses 5000 AES iterations, but the expanded-trace sentence says 200 iterations. Keep the correct number in both places and use one term, either iterations or executions.
\end{suggested}

\ifusecompiledplots
\begin{figure}[t]
\centering
\includegraphics[width=\linewidth]{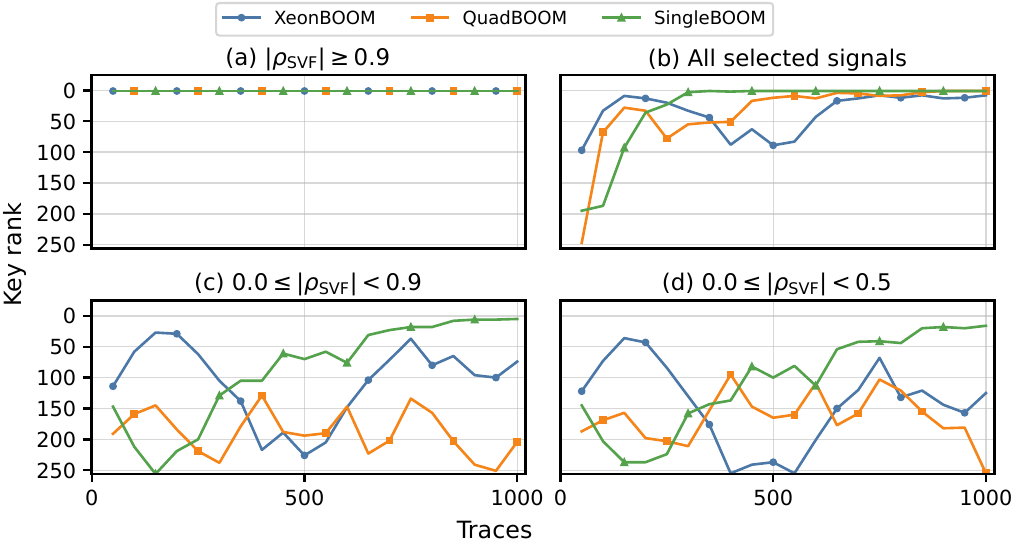}
\caption{\textbf{CPA attacks on SVF-based signal groups for XeonBOOM,
QuadBOOM, and SingleBOOM CPU configurations.} Lower rank top is better. Using only high-SVF signals
(Subfigure~(a)) recovers the correct key for all three designs within 50
traces, establishing a lower bound in simulation. Removing these signals
(Subfigures~(c) and~(d)) prevents recovery for QuadBOOM and XeonBOOM and
degrades SingleBOOM, highlighting their importance. Subfigure~(b) includes
both secret-key-related and background signal activity and therefore requires
more traces, showing that unrelated activity increases the attack effort.}
\label{fig:bit-hw-hw-svf-rank}
\end{figure}
\fi




\begin{table}[t]
\centering
\small
\setlength{\tabcolsep}{4pt}
\caption{Number of signals selected for
each SVF-correlation range in Figure~\ref{fig:bit-hw-hw-svf-rank}.}
\label{tab:svf-selected-signal-counts}
\begin{tabular}{@{}lrrr@{}}
\toprule
SVF selection &
\multicolumn{1}{c}{XeonBOOM} &
\multicolumn{1}{c}{QuadBOOM} &
\multicolumn{1}{c}{SingleBOOM} \\
\midrule
$|\rho_{\mathrm{SVF}}| \geq 0.9$
    & 71 & 68 & 68 \\
All selected signals
    & 174{,}499 & 66{,}180 & 43{,}866 \\ \midrule
$0.0 \leq |\rho_{\mathrm{SVF}}| < 0.9$
    & 174{,}428 & 66{,}112 & 43{,}798 \\
$0.0 \leq |\rho_{\mathrm{SVF}}| < 0.5$
    & 174{,}374 & 66{,}097 & 43{,}776 \\ \midrule
$0.0 \leq |\rho_{\mathrm{SVF}}| < 0.1$
    & 173{,}978 & 65{,}509 & 43{,}452 \\
\bottomrule
\end{tabular}
\end{table}

\guideblock{
\begin{itemize}
  \item Partition signals based on the maximum absolute correlation metric $|\rho|$.
  \item Construct three grouping strategies:
  \begin{itemize}
    \item Adjacent SVF intervals $[a,b)$.
    \item Lower-prefix cumulative groups $[0,\tau)$: $[0,1] \rightarrow [0,0.9) \rightarrow [0,0.8) \rightarrow \dots \rightarrow [0,0.1)$.
    \item Threshold-based groups $|\rho| \geq \tau$: $|\rho| \geq 0.9 \rightarrow |\rho| \geq 0.8 \rightarrow \dots \rightarrow |\rho| \geq 0$.
  \end{itemize}
  \item Perform CPA independently on each group.
  \item Measure attack effectiveness using MtD (minimum traces to reach rank~1).
  \item Compare trends across grouping strategies.
  \item Objective:
  \begin{itemize}
    \item Determine whether low-SVF signals leak useful information (individually or cumulatively).
    \item Quantify the dominance of high-SVF signals in key recovery.
    \item Characterize how leakage is distributed across the design.
    \item Evaluate how inclusion of low-SVF signals impacts attack efficiency.
  \end{itemize}
\end{itemize}
}

\begin{suggested}
\paragraph{Suggested text.} These simulation results justify the \name partitioning rule. High-SVF signals carry the dominant AES-correlated activity, while lower-SVF signals mainly add unrelated switching. \name therefore keeps high-SVF logic in the measured partition and moves lower-relevance activity away from the sensor when constraints allow it.
\end{suggested}

\begin{figure}[t]
    \centering
    \includegraphics[width=0.75\linewidth]{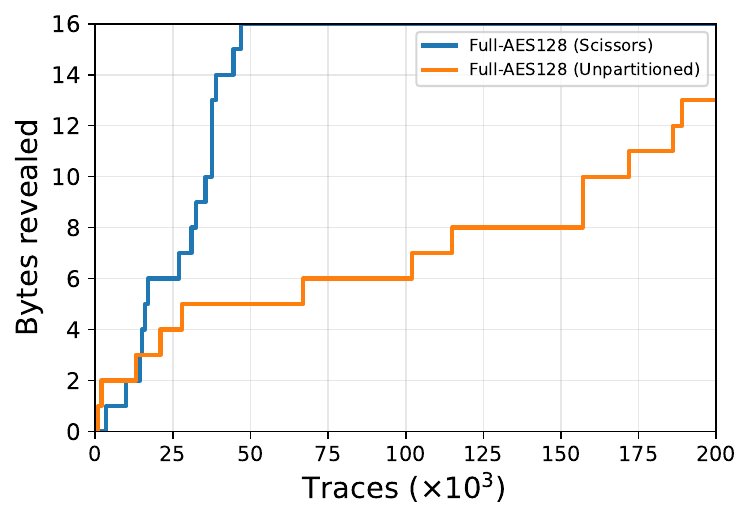}
   \caption{\textbf{Byte-reveal progress for Full-AES128.} We compare the same
   Full-AES128 design\protect\footnotemark{} in partitioned (\name) and
   unpartitioned forms under CPA. We consider a byte revealed at its first
   rank-1 trace, or at its first rank-2 trace if rank~1 never appears. \name
   reveals all 16 bytes by about 47{,}000 traces, while the unpartitioned form
   does not reveal all bytes within 200{,}000 traces.}
    \label{fig:byte-reveal}
\end{figure}
\footnotetext{The AES-128 reference design is from the Cryptographic Hardware IP repository: \url{https://github.com/ECSIS-lab/Cryptographic-Hardware-IPs}.}

\begin{figure}[t]
  \centering
  \includegraphics[width=\columnwidth]{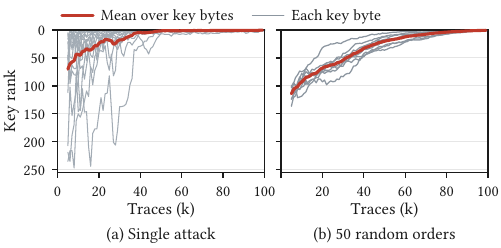}
  \caption{\logan{CPA on the HiFive P550. (a) Single attack on each of the 16 key bytes. (b) Mean of 50 attacks on each of key bytes 8 to 15, each using a random permutation of the 100{,}000 traces. Grey lines show individual key bytes and the red line their mean.}}
  \label{fig:sifive-keyrank}
\end{figure}




\ifusecompiledplots
\begin{figure}[t]
\centering
\includegraphics[width=\linewidth]{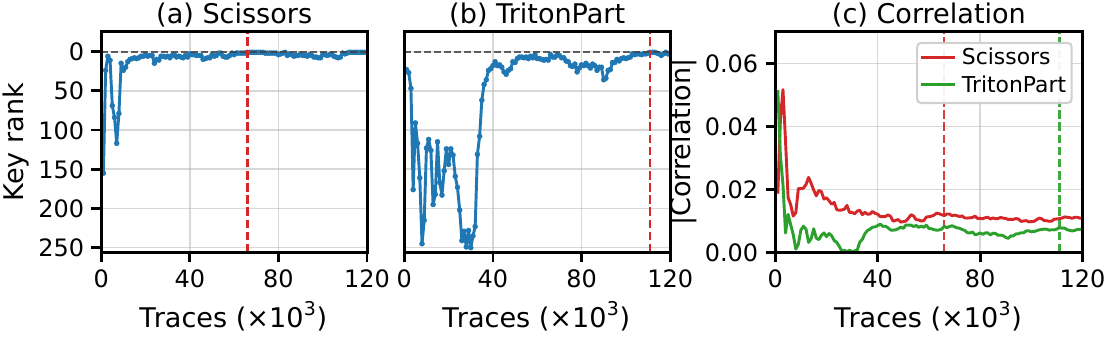}
\caption{\textbf{CPA attack comparison between \name and TritonPart for
QuadBOOM.} Subfigures~(a) and~(b) show frequency-domain CPA key-rank
progression for \name and TritonPart, respectively, while Subfigure~(c)
compares their absolute true-key correlations. Lower rank top is better; higher correlations are better.
QuadBOOM's four-core logic raises \name's disclosure cost from 12{,}000 traces
for SingleBOOM to 66{,}000 traces. This result remains a lower-noise bound for
an FPGA implementation because \name concentrates secret-dependent signals
while excluding unrelated logic. TritonPart not being secuirty-aware, inadvertently adds unrelated switching noise, and
requires 111{,}000 traces. \name achieves $1.48\times$ higher secret-key
correlation.}
\label{fig:quadboom-u200}
\end{figure}
\fi

\ifusecompiledplots
\begin{figure}[t]
\centering
\includegraphics[width=\linewidth]{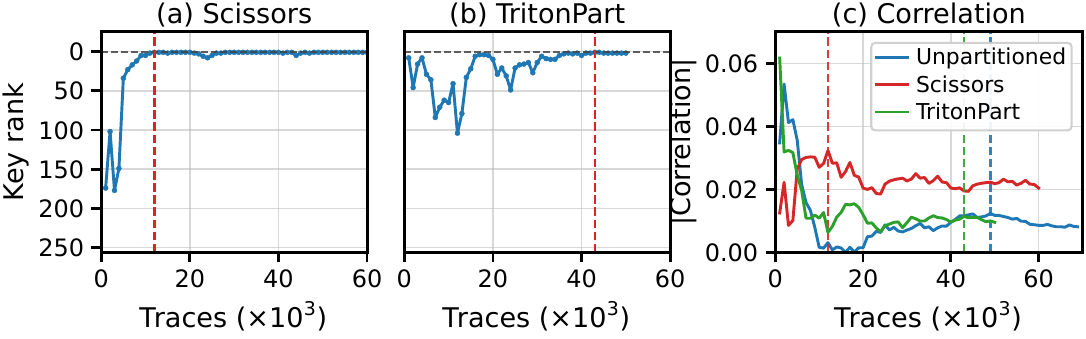}
\caption{\textbf{CPA attack comparison between \name and TritonPart for
SingleBOOM.} Subfigures~(a) and~(b) show frequency-domain CPA key rank
progression for \name and TritonPart, respectively, while Subfigure~(c)
compares their absolute true-key correlations. Lower key rank top is better; higher correlations are better. The
partitioned designs use two U200s. \name discloses the secret key with
12{,}000 traces, versus 49{,}000 for the unpartitioned design and 43{,}000 for
TritonPart, and improves correlation by $2.61\times$ and $2.83\times$,
respectively. By concentrating secret-dependent logic in the measured
partition, \name provides a lower-noise bound for an FPGA implementation.}
\label{fig:singleboom-cpa-comparison}
\end{figure}
\fi

\begin{table}[t]
\centering
\caption{Comparisons of CPA results for the SingleBOOM configuration for same attack hypothesis.}
\label{tab:d3-cpa-results}

\begin{minipage}[t]{0.49\columnwidth}
\vspace{0pt}
\centering
\scriptsize
\textbf{Attack results}\\[2pt]
\resizebox{\linewidth}{!}{%
\begin{tabular}{lrrr}
\toprule
\textbf{Design} & \textbf{U200s} & \textbf{Traces} & \textbf{$|\rho|$} \\
\midrule
Unpartitioned & 1 & 49k & 0.01240 \\
TritonPart    & 2 & 43k & 0.01144 \\
\name      & 2 & \textbf{12k} & \textbf{0.03236} \\
\bottomrule
\end{tabular}%
}
\end{minipage}
\hfill
\begin{minipage}[t]{0.49\columnwidth}
\vspace{0pt}
\centering
\scriptsize
\textbf{\name improvement}\\[2pt]
\resizebox{\linewidth}{!}{%
\begin{tabular}{lrr}
\toprule
\textbf{Baseline} & \textbf{Trace reduction} & \textbf{$|\rho|$ gain} \\
\midrule
Unpartitioned & \textbf{4.08$\times$} & \textbf{2.61$\times$} \\
TritonPart    & \textbf{3.58$\times$} & \textbf{2.83$\times$} \\
\bottomrule
\end{tabular}%
}
\end{minipage}
\end{table}

\begin{table}[t]
\centering
\caption{Attack performance for QuadBOOM partitioned across two U200 FPGAs. Traces denotes the first trace count at which the correct key reaches rank~1, and $|\rho|$ is the corresponding absolute correlation value.}
\label{tab:rocket64y4-attack-results}

\begin{minipage}[t]{0.49\columnwidth}
\vspace{0pt}
\centering
\scriptsize
\textbf{Attack results}\\[2pt]
\resizebox{\linewidth}{!}{%
\begin{tabular}{lrrr}
\toprule
\textbf{Design} & \textbf{U200s} & \textbf{Traces} & \textbf{$|\rho|$} \\
\midrule
TritonPart & 2 & 111k & 0.00788 \\
\name      & 2 & \textbf{66k} & \textbf{0.01167} \\
\bottomrule
\end{tabular}%
}
\end{minipage}
\hfill
\begin{minipage}[t]{0.49\columnwidth}
\vspace{0pt}
\centering
\scriptsize
\textbf{\name improvement}\\[2pt]
\resizebox{\linewidth}{!}{%
\begin{tabular}{lrr}
\toprule
\textbf{Baseline} & \textbf{Trace reduction} & \textbf{$|\rho|$ gain} \\
\midrule
TritonPart & \textbf{1.68$\times$} & \textbf{1.48$\times$} \\
\bottomrule
\end{tabular}%
}
\end{minipage}

\end{table}




\ifusecompiledplots
\begin{figure}[t]
    \centering
    \includegraphics[width=\linewidth]{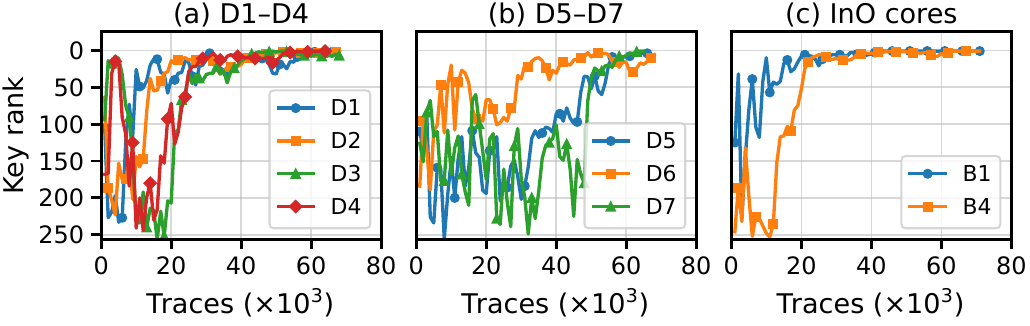}
    \caption{\textbf{CPA attack performance across processor configurations.}
    Table~\ref{tab:boom-configs} defines D1--D7. Subfigure~(a) considers
    D1--D4, which fit on the CW340, the largest available side-channel board.
    Subfigure~(b) considers D5--D7, which exceed the CW340's capacity, while
    Subfigure~(c) compares the B1 in-order single-core and B4 quad-core
    designs. Our automation allows us to evaluate this broad set of designs.
    Lower rank top is better. The
    simpler B1 and B4 designs reach rank~1 (recover the secret key) after
    approximately 40{,}000 traces. D1--D4 require around 60{,}000 traces,
    whereas D5--D7 require around 70{,}000 traces.}
    \label{fig:frequency-cpa-rank-comparison}
\end{figure}
\fi

\begin{suggested}
\paragraph{Suggested text.} The four-panel comparison should include all four completed panels or be reduced to the completed SiFive and U200 panels. If panels C and D represent baseline partitioning and \name, add those plots before making this figure central.
\end{suggested}

\label{subsec:sensitivity}
\guideblock{
\begin{itemize}
  \item  2--3 partition-cut variants (within vanilla and/or security-aware families).
\end{itemize}
}

\begin{suggested}
\paragraph{Suggested text.} Use the sensitivity study only after regenerating the plot from final data. The text should state what changes across CPU configurations, keep the workload and analysis pipeline fixed, and report whether microarchitectural complexity increases key-recovery trace count. Keep the platform metrics table only if every MtD, correlation, and SNR entry is filled.
\end{suggested}

\begin{suggested}
\paragraph{Suggested text.} Table~\ref{tab:rocket64y4_partition_method_comparison} motivates the per-instance localized-driver extraction used in the current \name flow. Moving complete instances preserves hierarchy but creates a wide serialized interface. Building one monolithic gate-level driver submodule reduces the interface, but expands partition-0 RTL substantially. The localized-driver method keeps the same SVF-guided intent while extracting only eligible local drivers inside user-approved instances. This gives a smaller interface than whole-instance extraction and avoids the RTL blow-up of the monolithic driver-cell partition.
\end{suggested}

\begin{table}[t]
\centering
\caption{Comparison of partition-boundary sizes produced by \name and
TritonPart for AES and PRESENT. TritonPart produces boundaries of 520 bits for AES and 307 bits for PRESENT.}
\label{tab:svf_threshold_partition_io_combined}
\resizebox{0.9\linewidth}{!}{%
\begin{tabular}{c rr rr}
\toprule
& \multicolumn{2}{c}{\textbf{AES}}
& \multicolumn{2}{c}{\textbf{PRESENT}} \\
\cmidrule(lr){2-3}\cmidrule(lr){4-5}
\textbf{\shortstack{SVF\\Threshold}}
& \textbf{\shortstack{Boundary\\Bits}}
& \textbf{\shortstack{Smaller than\\TritonPart}}
& \textbf{\shortstack{Boundary\\Bits}}
& \textbf{\shortstack{Smaller than\\TritonPart}} \\
\midrule
$\geq 0.9$ & 256 & $2.03\times$ & 256 & $1.20\times$ \\
\bottomrule
\end{tabular}%
}
\end{table}

\section{Measurement Evaluation}
\label{sec:evaluation}

\readytoreview{This section evaluates whether the secret-dependent activity identified in RTL simulation remains observable after FPGA implementation. We first establish a ASIC reference using the HiFive P550 platform, then evaluate \name on SingleBOOM and QuadBOOM using a two-U200 FPGA testbed, and finally extend the study to hardware AES and PRESENT accelerators. As Table~\ref{tab:boom-configs} shows, QuadBOOM closely resembles the P550 configuration~\cite{lam2025p550,lam2025riscvprogress}. Moreover, QuadBOOM exceeds the capacity of the CW340, the largest single-FPGA side-channel platform available, making it a representative candidate for multi-FPGA evaluation. Across these platforms, we compare measurements to disclosure (MtD) and secret-key correlation under the same Hamming-weight leakage model.


\textbf{HiFive P550 Golden Reference.} We collect 100{,}000 power traces from unprotected AES-128 running on a HiFive
P550 under Ubuntu 24.04. GPIO triggers identify the S-box window, and a Keysight
oscilloscope samples the 0.8-V CPU supply rail at 1~GS/s. Because the heatsink
is removed for supply access, the processor operates at 500~MHz rather than
its 1.4-GHz peak frequency.

Figure~\ref{fig:sifive-keyrank}(a) shows that most key bytes reach rank~1
within approximately 60{,}000 traces and all 16 bytes do so within 100{,}000
traces. To remove any bias in the trace order, we also perform 50 randomized
CPA attacks for each byte 8--15. Across these attacks, the mean
correct-key rank is 34, 13, and 5 at 40{,}000, 60{,}000, and 80{,}000 traces,
respectively (Figure~\ref{fig:sifive-keyrank}(b)). Full-key recovery
requires approximately 100{,}000 traces. Successful key recovery confirms that the AES S-box Hamming-weight hypothesis used throughout our evaluation exposes exploitable leakage on a commercial OoO processor under real supply-rail and board noise. The subsequent FPGA experiments evaluate the same leakage hypothesis before silicon, while comparisons among the unpartitioned, TritonPart, and \name implementations isolate the effect of partitioning using the same RTL and testbed. For all the subsequent CPU evaluations using FPGA measurements, we only show 1 byte attacks.

\pehcomment{{\color{red} \bf Peh: Explain how this golden reference is used to validate subsequent results.} {\color{purple} yash: updated the above text.}}


\textbf{SingleBOOM.} Figure~\ref{fig:singleboom-cpa-comparison} and
Table~\ref{tab:d3-cpa-results} compare the unpartitioned, TritonPart, and
\name implementations of SingleBOOM. \name reaches rank~1 using
$4.08\times$ fewer traces than the unpartitioned design and $3.58\times$
fewer traces than TritonPart. We also observe that \name achieves better
secret-key correlation: $2.61\times$ and $2.83\times$ higher than the
unpartitioned design and TritonPart, respectively. In noiseless simulation,
the correct key reaches rank~1 with only 300 traces. Physical
measurements require more traces because of various physical noise sources
and unrelated switching. The measured results show that \name preserves the
identified leakage better than unpartitioned, and TritonPart.


\textbf{QuadBOOM.} QuadBOOM does not fit on the CW340, the largest single-FPGA side-channel
platform available in our setup, making multi-FPGA partitioning unavoidable.
In our evaluation, TritonPart produces a symmetric split: cores~0 and~3,
together with their private L1 caches, are mapped to measured partition, while cores~1
and~2 and their L1 caches are mapped to support partition. \pehcomment{{\color{red} \bf Peh: harts? cores? parts?}} However, TritonPart is not
security-aware and does not optimize which leakage-relevant logic is placed
in the measured FPGA. In contrast, \name explicitly maps such logic to the
measured partition.

As shown in Table~\ref{tab:rocket64y4-attack-results}, \name requires
66{,}000 traces to reach rank~1, compared with 111{,}000 traces for
TritonPart, reducing MtD by $1.68\times$. We also observe that \name achieves
$1.48\times$ higher correct-key correlation. For context, noiseless QuadBOOM
RTL simulation reaches rank~1 with 800 traces, whereas full-key recovery on
the physical HiFive P550 requires approximately 100{,}000 traces. \name's lower trace requirement represents an empirical
lower bound because it preserves the same RTL while
excluding much of the unrelated switching from the measured partition. Overall, these results illustrate the gap between idealized simulation and
physical measurement, while the QuadBOOM comparison shows that
security-aware partitioning requires fewer traces than the baseline partitioning method.

\pehcomment{{\color{red} \bf Peh: Scissors require fewer traces than the golden reference board? Yet it's still accurate? Explain. {\color{purple} yash: added one sentence to the above text to added this. } }}


\textbf{Beyond CPUs}. We also apply \name to dedicated AES-128 and PRESENT accelerators. For AES,
\name reduces the required traces by more than $4\times$ and recovers all
16 key bytes, whereas the unpartitioned design recovers 13. Our flow does not constrain sensor proximity to high-SVF logic, so we sweep
sensor placement to verify that recovery does not depend on a favorable
location. Repeating the attack at four locations, including the default sensor location~\cite{rds2023} and one location in each of the U200's three Super Logic Regions (SLR0, SLR1, and SLR2), reveals the key at every location, showing that the result does not depend on one favorable placement. We use AES accelerator because each placement builds in about 5\,min, making it more practical than QuadBOOM for a similar sweep (Table~\ref{tab:cpu-partition-cost}).  \pehcomment{{\color{red} \bf Peh: Explain why sensor location is explored only for the accelerators, and not for the CPUs? Seems orthogonal}} \name also
outperforms TritonPart for both AES and PRESENT (Table~\ref{tab:svf_threshold_partition_io_combined}), demonstrating that the method
generalizes beyond processor workloads.

Across the P550, SingleBOOM, QuadBOOM, and hardware accelerators, the measured
results confirm that RTL-visible leakage survives physical implementation.
More importantly, \name consistently improves both rank convergence and
correct-key correlation over baseline partitioning method.}

\section{Implementation Cost and Scalability}
\label{sec:overhead}

\readytoreview{This section evaluates the implementation cost of obtaining the preceding
measurement improvements. We consider RTL-to-bitstream time, SVF-analysis and partitioning time, partition-interface width, FPGA utilization, and scalability to larger out-of-order cores.

\textbf{Partition-Generation Cost.} TritonPart requires a synthesized netlist for graph partitioning. In our flow,
Yosys first maps the design to an ASIC standard-cell library, after which
TritonPart partitions the resulting gate-level netlist. Although these
partitions preserve the required functionality, they are optimized for an
ASIC technology rather than an FPGA. Vivado must therefore resynthesize and
remap each partition to U200 resources. This additional mapping step is
unavoidable when using the TritonPart output in the FPGA evaluation.
In contrast, \name performs workload-guided SVF analysis and moves
high-SVF RTL instances between partitions. The resulting partitions retain
their RTL representation and can enter the FPGA synthesis flow directly.
SVF analysis takes 318.89~s for QuadBOOM and 1,506.39~s for XeonBOOM. For
XeonBOOM, the complete \name preprocessing flow takes 2\,h~59\,min, compared
with 16\,h~48\,min for the Yosys--TritonPart flow, a $5.61\times$ speedup.



\textbf{FPGA Implementation Cost.} Table~\ref{tab:cpu-partition-cost} reports the logical partition interface,
final placed CLB LUT utilization, and complete RTL-to-bitstream time for the QuadBOOM implementation. }

\begin{table}[t]
\centering
\caption{Implementation cost for QuadBOOM.}
\label{tab:cpu-partition-cost}

\resizebox{\columnwidth}{!}{%
\begin{tabular}{lrrrr}
\toprule
\textbf{Partitioning} &
\textbf{\shortstack{Interface bits\\F1$\rightarrow$F2 /
F2$\rightarrow$F1 (total)}} &
\textbf{\shortstack{FPGA~1\\CLB LUTs}} &
\textbf{\shortstack{FPGA~2\\CLB LUTs}} &
\textbf{\shortstack{Build time\\(h:mm)}} \\
\midrule
TritonPart
  & 273 / 487 (760)
  & 694,707 (58.76\%)
  & 620,610 (52.49\%)
  & 61:39 \\
\name
  & 4,013 / 2,397 (6,410)
  & 919,795 (77.80\%)
  & 75,593 (6.39\%)
  & \textbf{6:40} \\
\bottomrule
\end{tabular}%
}
\end{table}

\readytoreview{The cost of preserving leakage-relevant RTL is a wider logical interface:
\name uses 6,410 interface bits, compared with 760 for TritonPart. TritonPart
places two cores and their L1 caches on FPGA~2, resulting in 620,610 CLB LUTs
on FPGA~2 and 694,707 on FPGA~1. \name instead keeps most of the processor on
FPGA~1 and moves only the five selected high-SVF instances to FPGA~2.
Consequently, FPGA~2 uses only 75,593 CLB LUTs, while FPGA~1 uses 919,795.

The ASIC-netlist remapping required by TritonPart also increases Vivado build time. By preserving RTL throughout the FPGA flow, \name reduces complete RTL-to-bitstream generation from 61\,h~39\,min to 6\,h~40\,min, a $9.25\times$ speedup. \name also places only leakage-relevant instances on the measured FPGA, substantially reducing FPGA~2 utilization, although the resulting partition interface is wider. In our implementation, this wider interface makes \name's overall
emulation slower. However, \name dramatically reduces the number of traces\footnote{Each trace contains 60 samples of 128 bits each, totaling 960 bytes. At the measured UART throughput of 7.5~kB/s, transferring one trace takes 0.128~s.}
required, making the overall attack faster. For a single-byte attack on
QuadBOOM, \name takes 2.35\,h, whereas TritonPart takes 3.95\,h, making
\name $1.68\times$ faster. We expect the attack time to scale proportionally with the number of key bytes. Thus, a 16-byte attack takes approximately 37.56\,h with \name and 63.16\,h with TritonPart, reducing the wall-clock attack time by approximately 25.60\,h.}


\pehcomment{Peh: What about emulation time to run one trace? That's slower on Scissors due to the comm overhead vs TritonPart? Though the number of traces that need to be run is dramatically reduced, and hence overall attack time is faster. {\color{purple} yash: added the overall attack time and a projection for 16 byte attack.}}

\section{Related Work}
\label{sec:related}
\readytoreview{


\textbf{Side-channel attacks on complex processors.} Prior work has shown that modern processors leak through both telemetry-based and probe-based measurements. Hertzbleed~\cite{wang2022hertzbleed} exploits frequency-scaling effects to convert power leakage into timing leakage. PLATYPUS~\cite{lipp2021platypus} and Collide+Power~\cite{kogler2023collide+} use software-accessible power telemetry to observe processor-level leakage. Other work uses physical power or EM measurements on real processors, including Apple vs. EMA~\cite{haas2022apple}, SpectrEM~\cite{de2023spectrem}, and Reality Check on Side-Channels~\cite{boyapally2025reality}. These studies establish that complex processors can leak secrets during real execution, motivating pre-silicon frameworks that can evaluate such leakage before fabrication and at scales beyond single-FPGA emulation.

FPGA-based processor studies provide a pre-silicon alternative. Desmet et al.~\cite{desmet2026masking} demonstrate that OoO execution can break assumptions behind software masking, but their study uses a simplified OoO RISC-V processor. CryptRISC~\cite{srivastava2026cryptrisc} studies an in-order RISC-V processor with hardware support for protected cryptographic execution. Modern OoO designs have larger ROBs, physical register files, issue queues, and caches, which increase both leakage complexity and FPGA resource demand~\cite{chen2025paradise}. \name targets this underexplored setting by supporting larger OoO designs with multi-FPGA emulation and treating partitioning as a factor that can improve leakage observability.


\textbf{Multi-FPGA emulation and partitioning techniques.} Prior multi-FPGA partitioning work primarily optimizes capacity, timing, and communication. TritonPart performs constraint-driven graph partitioning for physical design~\cite{bustany2023open}. FireAxe uses user-guided module extraction to partition large BOOM designs across FPGAs~\cite{whangbo2024fireaxe}. Earlier processor prototypes, such as Intel Nehalem FPGA emulation, use microarchitectural clustering to manage inter-FPGA communication~\cite{schelle2010intel}. These approaches improve emulation scalability, but they do not consider side-channel leakage during partitioning.

\name adds leakage guidance to multi-FPGA partitioning. It uses SVF scores to place key-correlated signals in the measured FPGA partition and evaluates both the side-channel benefit and the implementation cost.}

\section{Discussion}
\label{sec:discussion}
\readytoreview{


\textbf{Generality and Applicability.} \name preserves the original RTL behavior because it only redistributes logic across FPGA partitions, while taking into account leakage of security data. Given a workload, a leakage hypothesis, and simulated activity, the same approach can apply beyond AES and OoO cores to other hardware. Further, by leveraging an on-chip sensor~\cite{rds2023}, the framework eliminates the need for an external oscilloscope. Automation allowed us to evaluate a wide range of CPU configurations (see Figure~\ref{fig:frequency-cpa-rank-comparison}) using physically measured traces. We plan to
open-source the framework to support further evaluation of other workloads,
leakage hypotheses. 

\pehcomment{Peh: Ok to mention other FPGA boards here? :) {\color{purple} yash: dropped other boards.} } 


\textbf{Limitations.} The approach supports pre-silicon evaluation but does not replace final silicon measurement. FPGA LUTs, routing, and power delivery differ from ASIC implementations, so absolute trace counts do not transfer directly. Fast ASIC simulation that captures realistic measurement noise remains an open problem.}

\section{Conclusion}
\label{sec:conclusion}

This paper presents \name, an SVF-guided multi-FPGA partitioning flow for pre-silicon power side-channel evaluation of complex OoO processors. \name identifies key-correlated signals, extracts the corresponding logic, and maps it to the measured FPGA to improve leakage observability during CPA. Our results show that partitioning is not only an implementation mechanism for fitting large designs across FPGAs, but also a security-relevant design choice that affects trace quality and attack performance. Across the physical FPGA evaluations, \name reduces MtD by up to $4.08\times$,
improves correct-key correlation by up to $2.83\times$, and reduces QuadBOOM
RTL-to-bitstream generation time by $9.25\times$ relative to the corresponding
baselines.


\begin{suggested}
\paragraph{Suggested text.} This paper presents \name, an SVF-guided multi-FPGA partitioning flow for pre-silicon power side-channel evaluation of complex OoO processors. \name identifies key-correlated switching, extracts the corresponding logic, and maps it to the measured FPGA so CPA observes cleaner leakage. The final conclusion should insert measured trace-count reduction and overhead numbers, then close with the main lesson: in multi-FPGA side-channel evaluation, partitioning is both an implementation step and a security-relevant control knob.
\end{suggested}

\clearpage
\bibliographystyle{ACM-Reference-Format}
\bibliography{refs}

@inproceedings{mateos2010new,
  title={A new correlation frequency analysis of the side channel},
  author={Mateos, Edgar and Gebotys, Catherine H},
  booktitle={Proceedings of the 5th Workshop on Embedded Systems Security},
  pages={1--8},
  year={2010}
}

@inproceedings{veshchikov2017use,
  title={Use of simulators for side-channel analysis},
  author={Veshchikov, Nikita and Guilley, Sylvain},
  booktitle={2017 IEEE European Symposium on Security and Privacy Workshops (EuroS\&PW)},
  pages={104--112},
  year={2017},
  organization={IEEE}
}

@inproceedings{tong2024easypart,
  title={Easypart: An effective and comprehensive hypergraph partitioner for fpga-based emulation},
  author={Tong, Shengbo and Li, Haoyuan and Xu, Jiahao and Pei, Chunyan and Yu, Wenjian and Liu, Shengjun and Shen, Jian},
  booktitle={Proceedings of the 43rd IEEE/ACM International Conference on Computer-Aided Design},
  pages={1--9},
  year={2024}
}

@article{bobda2022future,
  title={The future of FPGA acceleration in datacenters and the cloud},
  author={Bobda, Christophe and Mbongue, Joel Mandebi and Chow, Paul and Ewais, Mohammad and Tarafdar, Naif and Vega, Juan Camilo and Eguro, Ken and Koch, Dirk and Handagala, Suranga and Leeser, Miriam and others},
  journal={ACM Transactions on Reconfigurable Technology and Systems (TRETS)},
  volume={15},
  number={3},
  pages={1--42},
  year={2022},
  publisher={ACM New York, NY}
}

@article{tessier2015reconfigurable,
  title={Reconfigurable computing architectures},
  author={Tessier, Russell and Pocek, Kenneth and DeHon, Andre},
  journal={Proceedings of the IEEE},
  volume={103},
  number={3},
  pages={332--354},
  year={2015},
  publisher={IEEE}
}

@inproceedings{kapre2009accelerating,
  title={Accelerating SPICE model-evaluation using FPGAs},
  author={Kapre, Nachiket and DeHon, Andr{\'e}},
  booktitle={2009 17th IEEE Symposium on Field Programmable Custom Computing Machines},
  pages={37--44},
  year={2009},
  organization={IEEE}
}

@inproceedings{dai2017foregraph,
  title={ForeGraph: Exploring large-scale graph processing on multi-FPGA architecture},
  author={Dai, Guohao and Huang, Tianhao and Chi, Yuze and Xu, Ningyi and Wang, Yu and Yang, Huazhong},
  booktitle={Proceedings of the 2017 ACM/SIGDA International Symposium on Field-Programmable Gate Arrays},
  pages={217--226},
  year={2017}
}

@inproceedings{zha2020virtualizing,
  title={Virtualizing FPGAs in the cloud},
  author={Zha, Yue and Li, Jing},
  booktitle={Proceedings of the Twenty-Fifth International Conference on Architectural Support for Programming Languages and Operating Systems},
  pages={845--858},
  year={2020}
}

@inproceedings{mazraeli2023partitioning,
  title={Partitioning large-scale, multi-FPGA applications for the data center},
  author={Mazraeli, Mohammadmahdi and Gao, Yu and Chow, Paul},
  booktitle={2023 33rd International Conference on Field-Programmable Logic and Applications (FPL)},
  pages={253--258},
  year={2023},
  organization={IEEE}
}

@inproceedings{tian2021remote,
  title={Remote power attacks on the versatile tensor accelerator in multi-tenant FPGAs},
  author={Tian, Shanquan and Moini, Shayan and Wolnikowski, Adam and Holcomb, Daniel and Tessier, Russell and Szefer, Jakub},
  booktitle={2021 IEEE 29th Annual International Symposium on Field-Programmable Custom Computing Machines (FCCM)},
  pages={242--246},
  year={2021},
  organization={IEEE}
}

@article{schlag2023high,
  title={High-quality hypergraph partitioning},
  author={Schlag, Sebastian and Heuer, Tobias and Gottesb{\"u}ren, Lars and Akhremtsev, Yaroslav and Schulz, Christian and Sanders, Peter},
  journal={ACM Journal of Experimental Algorithmics},
  volume={27},
  pages={1--39},
  year={2023},
  publisher={ACM New York, NY}
}

@inproceedings{karypis1997multilevel,
  title={Multilevel hypergraph partitioning: Application in VLSI domain},
  author={Karypis, George and Aggarwal, Rajat and Kumar, Vipin and Shekhar, Shashi},
  booktitle={Proceedings of the 34th annual Design Automation Conference},
  pages={526--529},
  year={1997}
}

@inproceedings{zheng2021topopart,
  title={Topopart: a multi-level topology-driven partitioning framework for multi-fpga systems},
  author={Zheng, Dan and Zang, Xinshi and Wong, Martin DF},
  booktitle={2021 IEEE/ACM International Conference On Computer Aided Design (ICCAD)},
  pages={1--8},
  year={2021},
  organization={IEEE}
}

@misc{keysight_msox3104t,
  author       = {{Keysight Technologies}},
  title        = {{MSOX3104T} Mixed Signal Oscilloscope: 1 {GHz}, 4 Analog Plus 16 Digital Channels},
  howpublished = {\url{https://www.keysight.com/sg/en/product/MSOX3104T/mixed-signal-oscilloscope-1-ghz-4-analog-16-digital-channels.html}},
  note         = {Accessed: 2026-10-02}
}

@misc{caltest_ct4200_series,
  author       = {{Cal Test Electronics}},
  title        = {{CT4200} Series Datasheet: Passive Voltage Oscilloscope Probes},
  year         = {2018},
  howpublished = {\url{https://as.caltestelectronics.com/public/attachments/CT4200_Series_datasheet.pdf}},
  note         = {Accessed: 2026-10-02}
}

@article{kim2016strober,
  title={Strober: Fast and accurate sample-based energy simulation for arbitrary RTL},
  author={Kim, Donggyu and Izraelevitz, Adam and Celio, Christopher and Kim, Hokeun and Zimmer, Brian and Lee, Yunsup and Bachrach, Jonathan and Asanovi{\'c}, Krste},
  journal={ACM SIGARCH Computer Architecture News},
  volume={44},
  number={3},
  pages={128--139},
  year={2016},
  publisher={ACM New York, NY, USA}
}

@inproceedings{guntur2014side,
  title={Side-channel attack user reference architecture board SAKURA-G},
  author={Guntur, Hendra and Ishii, Jun and Satoh, Akashi},
  booktitle={2014 IEEE 3rd global conference on consumer electronics (GCCE)},
  pages={271--274},
  year={2014},
  organization={IEEE}
}

@inproceedings{meynard2010characterization,
  title={Characterization of the electromagnetic side channel in frequency domain},
  author={Meynard, Olivier and R{\'e}al, Denis and Guilley, Sylvain and Flament, Florent and Danger, Jean-Luc and Valette, Fr{\'e}d{\'e}ric},
  booktitle={International Conference on Information Security and Cryptology},
  pages={471--486},
  year={2010},
  organization={Springer}
}

@misc{lam2025p550,
  author       = {Lam, Chester},
  title        = {Inside {SiFive}'s {P550} Microarchitecture},
  year         = {2025},
  month        = jan,
  howpublished = {\url{https://chipsandcheese.com/p/inside-sifives-p550-microarchitecture}},
  note         = {Accessed: 2026-09-30}
}

@misc{lam2025riscvprogress,
  author       = {Lam, Chester},
  title        = {A {RISC-V} Progress Check: Benchmarking {P550} and {C910}},
  year         = {2025},
  month        = jan,
  howpublished = {\url{https://chipsandcheese.com/p/a-risc-v-progress-check-benchmarking}},
  note         = {Accessed: 2026-09-30}
}

@misc{amd_ultrascale_fpga_selection_guide_2016,
  author       = {{Advanced Micro Devices, Inc.}},
  title        = {{UltraScale FPGA Product Tables and Product Selection Guide (XMP102)}},
  year         = {2016},
  howpublished = {\url{https://docs.amd.com/v/u/en-US/ultrascale-fpga-product-selection-guide}},
  note         = {Document ID: XMP102; Revision: 1.7. Accessed: 2026-09-13}
}

@inproceedings{balasch2015dpa,
  title={DPA, bitslicing and masking at 1 GHz},
  author={Balasch, Josep and Gierlichs, Benedikt and Reparaz, Oscar and Verbauwhede, Ingrid},
  booktitle={International Workshop on Cryptographic Hardware and Embedded Systems},
  pages={599--619},
  year={2015},
  organization={Springer}
}

@inproceedings{wang2022hertzbleed,
  title={Hertzbleed: Turning power $\{$Side-Channel$\}$ attacks into remote timing attacks on x86},
  author={Wang, Yingchen and Paccagnella, Riccardo and He, Elizabeth Tang and Shacham, Hovav and Fletcher, Christopher W and Kohlbrenner, David},
  booktitle={31st USENIX Security Symposium (USENIX Security 22)},
  pages={679--697},
  year={2022}
}

@inproceedings{lipp2021platypus,
  title={PLATYPUS: Software-based power side-channel attacks on x86},
  author={Lipp, Moritz and Kogler, Andreas and Oswald, David and Schwarz, Michael and Easdon, Catherine and Canella, Claudio and Gruss, Daniel},
  booktitle={2021 IEEE Symposium on Security and Privacy (SP)},
  pages={355--371},
  year={2021},
  organization={IEEE}
}

@inproceedings{kogler2023collide+,
  title={$\{$Collide+ Power$\}$: Leaking inaccessible data with software-based power side channels},
  author={Kogler, Andreas and Juffinger, Jonas and Giner, Lukas and Gerlach, Lukas and Schwarzl, Martin and Schwarz, Michael and Gruss, Daniel and Mangard, Stefan},
  booktitle={32nd USENIX security symposium (USENIX Security 23)},
  pages={7285--7302},
  year={2023}
}

@inproceedings{boyapally2025reality,
  title = {Reality Check on {Side-Channels}: Lessons Learnt from Breaking {AES} on {ARM} {Cortex-A72} Processor with {Out-of-Order} Execution},
  author={Boyapally, Harishma and Jap, Dirmanto and Wu, Qianmei and Zhang, Fan and Bhasin, Shivam},
  booktitle={19th USENIX WOOT Conference on Offensive Technologies (WOOT 25)},
  pages={179--189},
  year={2025}
}

@article{desmet2026masking,
  title={Masking Out of Order: Side-Channel Leaks from Software-Masked Cryptography on Out-of-Order Processors},
  author={Desmet, Eden and Kundu, Suparna and Verbauwhede, Ingrid},
  journal={Cryptology ePrint Archive},
  year={2026}
}

@article{srivastava2026cryptrisc,
  title={CryptRISC: A Secure RISC-V Processor for High-Performance Cryptography with Power Side-Channel Protection},
  author={Srivastava, Amisha and Porwal, Muskan and Basu, Kanad},
  journal={arXiv preprint arXiv:2602.20285},
  year={2026}
}

@article{chen2025paradise,
  title={PARADISE: Criticality-aware instruction reordering for power attack resistance},
  author={Chen, Yun and Hajiabadi, Ali and Poussier, Romain and Tavva, Yaswanth and Diavastos, Andreas and Bhasin, Shivam and Carlson, Trevor E},
  journal={ACM Transactions on Architecture and Code Optimization},
  volume={22},
  number={1},
  pages={1--25},
  year={2025},
  publisher={ACM New York, NY}
}

@inproceedings{bustany2023open,
  title={An open-source constraints-driven general partitioning multi-tool for VLSI physical design},
  author={Bustany, Ismail and Gasparyan, Grigor and Kahng, Andrew B and Koutis, Ioannis and Pramanik, Bodhisatta and Wang, Zhiang},
  booktitle={2023 IEEE/ACM International Conference on Computer Aided Design (ICCAD)},
  pages={1--9},
  year={2023},
  organization={IEEE}
}

@inproceedings{whangbo2024fireaxe,
  title={FireAxe: partitioned FPGA-accelerated simulation of large-scale RTL designs},
  author={Whangbo, Joonho and Lim, Edwin and Zhang, Chengyi Lux and Anderson, Kevin and Gonzalez, Abraham and Gupta, Raghav and Krishnakumar, Nivedha and Karandikar, Sagar and Nikoli{\'c}, Borivoje and Shao, Yakun Sophia and others},
  booktitle={2024 ACM/IEEE 51st Annual International Symposium on Computer Architecture (ISCA)},
  pages={501--515},
  year={2024},
  organization={IEEE}
}

@article{anik2024resiliency,
  title={On the resiliency of protected masked s-boxes against template attack in the presence of temperature and aging misalignments},
  author={Anik, Md Toufiq Hasan and Danger, Jean-Luc and Guilley, Sylvain and Karimi, Naghmeh},
  journal={IEEE Transactions on Very Large Scale Integration (VLSI) Systems},
  volume={32},
  number={5},
  pages={911--924},
  year={2024},
  publisher={IEEE}
}

@article{kiaei2021saidoyoki,
  title={Saidoyoki: Evaluating side-channel leakage in pre-and post-silicon setting.},
  author={Kiaei, Pantea and Liu, Zhenyuan and Eren, Ramazan Kaan and Yao, Yuan and Schaumont, Patrick},
  journal={IACR Cryptol. ePrint Arch.},
  volume={2021},
  pages={1235},
  year={2021}
}

@inproceedings{prakriya2024tapa,
  title={TAPA-CS: Enabling Scalable Accelerator Design on Distributed HBM-FPGAs},
  author={Prakriya, Neha and Chi, Yuze and Basalama, Suhail and Song, Linghao and Cong, Jason},
  booktitle={Proceedings of the 29th ACM International Conference on Architectural Support for Programming Languages and Operating Systems, Volume 3},
  pages={966--980},
  year={2024}
}

@misc{kropotov2026emixemulatingsinglefpgalimits,
      title={EMiX: Emulating Beyond Single-FPGA Limits}, 
      author={Alexander Kropotov and Miquel Moreto and Behzad Salami},
      year={2026},
      eprint={2604.27012},
      archivePrefix={arXiv},
      primaryClass={cs.AR},
      url={https://arxiv.org/abs/2604.27012}, 
}

@inproceedings{chirkov2023smappic,
  title={Smappic: Scalable multi-fpga architecture prototype platform in the cloud},
  author={Chirkov, Grigory and Wentzlaff, David},
  booktitle={Proceedings of the 28th ACM International Conference on Architectural Support for Programming Languages and Operating Systems, Volume 2},
  pages={733--746},
  year={2023}
}

@inproceedings{karandikar2018firesim,
  title={FireSim: FPGA-accelerated cycle-exact scale-out system simulation in the public cloud},
  author={Karandikar, Sagar and Mao, Howard and Kim, Donggyu and Biancolin, David and Amid, Alon and Lee, Dayeol and Pemberton, Nathan and Amaro, Emmanuel and Schmidt, Colin and Chopra, Aditya and others},
  booktitle={2018 ACM/IEEE 45th Annual International Symposium on Computer Architecture (ISCA)},
  pages={29--42},
  year={2018},
  organization={IEEE}
}

@inproceedings{zhang2025heteroproto,
  title={HeteroProto: Automated RTL-to-Bitstream Framework for Heterogeneous Multi-FPGA SoC Prototyping},
  author={Zhang, Congwu and Wang, Panyu and Wang, Yazhou and Chen, Mingyu and Bao, Yungang and Chang, Yisong and Zhang, Ke},
  booktitle={2025 International Conference on Field Programmable Technology (ICFPT)},
  pages={101--110},
  year={2025},
  organization={IEEE}
}

@misc{modernsoc2024debao,
      author = {Debao Wang and Yiwen Gao and Yongbin Zhou and Xian Huang},
      title = {Revisiting a Realistic {EM} Side-Channel Attack on a Complex Modern {SoC}},
      howpublished = {Cryptology {ePrint} Archive, Paper 2024/1322},
      year = {2024},
      url = {https://eprint.iacr.org/2024/1322}
}

@inproceedings{longo2015soc,
  title={SoC it to EM: electromagnetic side-channel attacks on a complex system-on-chip},
  author={Longo, Jake and De Mulder, Elke and Page, Dan and Tunstall, Michael},
  booktitle={International Workshop on Cryptographic Hardware and Embedded Systems},
  pages={620--640},
  year={2015},
  organization={Springer}
}

@inproceedings{brier2004correlation,
  title={Correlation power analysis with a leakage model},
  author={Brier, Eric and Clavier, Christophe and Olivier, Francis},
  booktitle={International workshop on cryptographic hardware and embedded systems},
  pages={16--29},
  year={2004},
  organization={Springer}
}

@inproceedings{schelle2010intel,
  title={Intel Nehalem processor core made FPGA synthesizable},
  author={Schelle, Graham and Collins, Jamison and Schuchman, Ethan and Wang, Perrry and Zou, Xiang and Chinya, Gautham and Plate, Ralf and Mattner, Thorsten and Olbrich, Franz and Hammarlund, Per and others},
  booktitle={Proceedings of the 18th annual ACM/SIGDA international symposium on Field programmable gate arrays},
  pages={3--12},
  year={2010}
}

@INPROCEEDINGS{svfisca2012,
  author={Demme, John and Martin, Robert and Waksman, Adam and Sethumadhavan, Simha},
  booktitle={2012 39th Annual International Symposium on Computer Architecture (ISCA)}, 
  title={Side-channel vulnerability factor: A metric for measuring information leakage}, 
  year={2012},
  volume={},
  number={},
  pages={106--117},
  doi={10.1109/ISCA.2012.6237010}}

@inproceedings{de2023spectrem,
  title={$\{$SpectrEM$\}$: Exploiting Electromagnetic Emanations During Transient Execution},
  author={De Meulemeester, Jesse and Purnal, Antoon and Wouters, Lennert and Beckers, Arthur and Verbauwhede, Ingrid},
  booktitle={32nd USENIX Security Symposium (USENIX Security 23)},
  pages={6293--6310},
  year={2023}
}

@inproceedings{haas2022apple,
  title={Apple vs. EMA: electromagnetic side channel attacks on apple CoreCrypto},
  author={Haas, Gregor and Aysu, Aydin},
  booktitle={Proceedings of the 59th ACM/IEEE Design Automation Conference},
  pages={247--252},
  year={2022}
}

@inproceedings{azouaoui2020systematic,
  title={A systematic appraisal of side channel evaluation strategies},
  author={Azouaoui, Melissa and Bellizia, Davide and Buhan, Ileana and Debande, Nicolas and Duval, S{\'e}bastien and Giraud, Christophe and Jaulmes, {\'E}liane and Koeune, Fran{\c{c}}ois and Oswald, Elisabeth and Standaert, Fran{\c{c}}ois-Xavier and others},
  booktitle={International Conference on Research in Security Standardisation},
  pages={46--66},
  year={2020},
  organization={Springer}
}

@misc{lam2024lioncove,
  author       = {Chester Lam},
  title        = {Lion Cove: Intel's P-Core Roars},
  year         = {2024},
  month        = sep,
  day          = {27},
  howpublished = {\url{https://chipsandcheese.com/p/lion-cove-intels-p-core-roars}},
  note         = {Accessed October 4, 2026}
}

@misc{johnson_firestorm,
  author       = {Dougall Johnson},
  title        = {Firestorm Overview},
  howpublished = {\url{https://dougallj.github.io/applecpu/firestorm.html}},
  note         = {Apple Microarchitecture Research; accessed October 4, 2026}
}

@misc{xilinx_aurora_alveo_vitis_2022,
  author       = {{AMD Xilinx}},
  title        = {Using Aurora IP in Alveo with Vitis Flow},
  year         = {2022},
  howpublished = {\url{https://xilinx.github.io/Vitis-Tutorials/2022-1/build/html/docs/Hardware_Acceleration/Design_Tutorials/08-alveo_aurora_kernel/README.html}},
  note         = {Vitis Tutorials 2022.1, accessed October 4, 2026}
}

@inproceedings{o2014chipwhisperer,
  title={Chipwhisperer: An open-source platform for hardware embedded security research},
  author={O’flynn, Colin and Chen, Zhizhang},
  booktitle={International Workshop on Constructive Side-Channel Analysis and Secure Design},
  pages={243--260},
  year={2014},
  organization={Springer}
}

@inproceedings{ferrufino2023fobos,
  title={FOBOS 3: An open-source platform for side-channel analysis and benchmarking},
  author={Ferrufino, Eduardo and Beckwith, Luke and Abdulgadir, Abubakr and Kaps, Jens-Peter},
  booktitle={Proceedings of the 2023 Workshop on Attacks and Solutions in Hardware Security},
  pages={5--14},
  year={2023}
}

@article{drewes2024turn,
  title={Turn on, tune in, and listen up: Maximizing side-channel recovery in cross-platform time-to-digital converters},
  author={Drewes, Colin and Sheaves, Tyler and Weng, Olivia and Ryan, Keegan and Hunter, Bill and McCarty, Christopher and Kastner, Ryan and Richmond, Dustin},
  journal={ACM Transactions on Reconfigurable Technology and Systems},
  volume={17},
  number={3},
  pages={1--30},
  year={2024},
  publisher={ACM New York, NY}
}

@article{schellenberg2021inside,
  title={An inside job: Remote power analysis attacks on FPGAs},
  author={Schellenberg, Falk and Gnad, Dennis RE and Moradi, Amir and Tahoori, Mehdi B},
  journal={IEEE Design \& Test},
  volume={38},
  number={3},
  pages={58--66},
  year={2021},
  publisher={IEEE}
}

@inproceedings{quere2026snoopypower,
  title={SnoopyPower! Remote Power Attacks on Cache and Coherence Paths},
  author={Qu{\'e}r{\'e}, Eliott and Real, Maria M{\'e}ndez and Palumbo, Alessandro and Rokicki, Thomas and Bossuet, Lilian and Salvador, Rub{\'e}n},
  booktitle={2026 IEEE International Symposium on Hardware Oriented Security and Trust (HOST)},
  pages={1--13},
  year={2026},
  organization={IEEE}
}

@inproceedings{aftabjahani2021special,
  title={Special session: Cad for hardware security-automation is key to adoption of solutions},
  author={Aftabjahani, Sohrab and Kastner, Ryan and Tehranipoor, Mark and Farahmandi, Farimah and Oberg, Jason and Nordstrom, Anders and Fern, Nicole and Althoff, Alric},
  booktitle={2021 IEEE 39th VLSI Test Symposium (VTS)},
  pages={1--10},
  year={2021},
  organization={IEEE}
}

@article{papagiannopoulos2023side,
  title={The side-channel metrics cheat sheet},
  author={Papagiannopoulos, Kostas and Glamo{\v{c}}anin, Ognjen and Azouaoui, Melissa and Ros, Dorian and Regazzoni, Francesco and Stojilovi{\'c}, Mirjana},
  journal={ACM Computing Surveys},
  volume={55},
  number={10},
  pages={1--38},
  year={2023},
  publisher={ACM New York, NY}
}

@inproceedings{zhao2018fpga,
  title={FPGA-based remote power side-channel attacks},
  author={Zhao, Mark and Suh, G Edward},
  booktitle={2018 IEEE symposium on security and privacy (SP)},
  pages={229--244},
  year={2018},
  organization={IEEE}
}

@misc{amd_alveo_u200,
  author       = {{Advanced Micro Devices, Inc.}},
  title        = {{AMD Alveo U200 Data Center Accelerator Card}},
  howpublished = {\url{https://www.amd.com/en/products/accelerators/alveo/u200/a-u200-a64g-pq-g.html}},
  note         = {Accessed: 2026-09-16}
}

@article{tavva2025ctscan,
  title={Ctscan: A cgra-based platform for the emulation of power side-channel attacks on edge cpus},
  author={Tavva, Yaswanth and Juneja, Rohan and Carlson, Trevor E and Peh, Li-Shiuan},
  journal={ACM Transactions on Reconfigurable Technology and Systems},
  volume={18},
  number={2},
  pages={1--36},
  year={2025},
  publisher={ACM New York, NY}
}

@article{rds2023, title={RDS: FPGA Routing Delay Sensors for Effective Remote Power Analysis Attacks}, volume={2023}, url={https://tches.iacr.org/index.php/TCHES/article/view/10293}, DOI={10.46586/tches.v2023.i2.543-567}, number={2}, journal={IACR Transactions on Cryptographic Hardware and Embedded Systems}, author={Spielmann, David and Glamočanin, Ognjen and Stojilović, Mirjana}, year={2023}, month={Mar.}, pages={543--567}}

@inproceedings{babb1993virtual,
  title={Virtual wires: Overcoming pin limitations in FPGA-based logic emulators},
  author={Babb, Jonathan and Tessier, Russell and Agarwal, Anant},
  booktitle={[1993] Proceedings IEEE Workshop on FPGAs for Custom Computing Machines},
  pages={142--151},
  year={1993},
  organization={IEEE}
}

@misc{cw310,
  author       = {NewAE Technology},
  title        = {CW310 Bergen Board (Kintex FPGA Target)},
  year         = {2025},
  url          = {https://rtfm.newae.com/Targets/CW310%20Bergen%20Board/},
  note         = {Accessed: 2025-02-09}
}

@misc{cw340,
  author       = {NewAE Technology},
  title        = {CW340 Luna Board (OpenTitan Baseboard)},
  year         = {2025},
  url          = {https://rtfm.newae.com/Targets/CW340%20Luna%20Board/},
  note         = {Accessed: 2025-02-09}
}

\end{document}